\documentclass[11pt,a4paper]{article}
\pdfoutput=1

\usepackage{jcappub}
\usepackage[colorlinks=true,linkcolor=blue,citecolor=blue,urlcolor=blue]{hyperref}

\usepackage[margin=0.95in]{geometry}

\usepackage[utf8]{inputenc}
\usepackage{graphicx}
\usepackage{verbatim}
\usepackage{epsfig}
\usepackage{subcaption} 
\usepackage{color}
\usepackage{multirow}
\usepackage{comment}
\usepackage{slashed}
\usepackage{amsmath}
\usepackage[normalem]{ulem}
\usepackage{framed}
\usepackage{float}
\usepackage{amssymb,amsmath,amsfonts,mathrsfs}
\usepackage[dvipsnames]{xcolor}
\usepackage{graphicx}
\usepackage{longtable}
\usepackage{verbatim}
\usepackage{color}
\usepackage[mathscr]{euscript}
\usepackage{framed}
\usepackage{mdframed}  

\usepackage{cancel}
\usepackage{color}
\usepackage{hyperref}
\usepackage{algorithm} 
\usepackage{algpseudocode} 
\usepackage{comment}
\hypersetup{colorlinks, citecolor=bluscuro, linkcolor=blue, urlcolor=bluscuro}
\definecolor{rossos}{cmyk}{0,1,1,0.55}
\definecolor{bluscuro}{rgb}{0.15, 0.2, .85}
\definecolor{bluchiaro}{cmyk}{1,.3,0.,0.1}

\numberwithin{equation}{section}

\usepackage{tikz}

\usepackage{tcolorbox}

\title{Smoluchowski Coagulation Equation and the Evolution of Primordial Black Hole Clusters}
\author[a]{Borui Zhang,}
\author[a]{Wei-Xiang Feng}
\author[a,b]{and Haipeng An}

\affiliation[a]{Department of Physics, Tsinghua University, Beijing 100084, China}
\affiliation[b]{Center for High Energy Physics, Tsinghua University, Beijing 100084, China}

\abstract{
In Ref.\;\cite{Zhang:2025tgm}, we demonstrate that the high-redshift supermassive black holes  in the so-called ``little red dots'' discovered by James Webb Space Telescope (JWST) can be explained by the primordial black hole (PBH) clustering on small scales. In this paper, we present a comprehensive simulation of the successive PBH mergers within a cluster by solving the Smoluchowski coagulation equation. We derive the coagulation kernel considering both cases with and without the effects of mass segregation. Then we employ the Monte Carlo method to solve the equation, implementing the full-conditioning scheme using the discrete inverse transformation method. Our simulations determine the runaway timescales of clusters and the mass population evolution of PBHs across a wide range of cosmic redshifts, depending on the number of PBHs within the cluster and the associated density. 
}
\emailAdd{wxfeng@mail.tsinghua.edu.cn}
\emailAdd{anhp@mail.tsinghua.edu.cn}
\begin{document}

\maketitle
\flushbottom

\section{Introduction}
Primordial black holes (PBHs) are associated with the physics of the early Universe, as their formation is tied directly to the density fluctuations during or right after the cosmic inflation. Unlike stellar black holes, PBHs could span a wide range of masses—from sub‑atomic scales up to millions of solar masses—making them powerful probes of high‑energy processes, phase transitions, and the amplitude of primordial density perturbations. They might serve as seeds for supermassive black holes (SMBHs) observed at high redshifts\;\cite{Duechting:2004dk,Kawasaki:2012kn,Ziparo:2024nwh,Zhang:2025tgm,Zhang:2025oyl}, influence the evolution of cosmic structures\;\cite{Carr:1974nx,Carr:1975qj,Dokuchaev:2008hz,Carr:2018rid,Inman:2019wvr,Liu:2022bvr,Su:2023jno,Huang:2024aog,Huang:2024koy,Dayal:2025aiv}, and potentially account for all or part of the dark matter\;\cite{Zeldovich:1967lct,Hawking:1971ei,Carr:2016drx,Carr:2020gox,Gorton:2024cdm,Dai:2024guo,Carr:2026hot}.

PBH mergers produce gravitational waves whose spectrum carries imprints of the underlying formation history, offering further observational tests with current and future detectors, see Ref.\;\cite{Carr:2026hot} for a recent review. In particular, the recent LIGO--Virgo--KAGRA (LVK) collaboration has detected hundreds of black hole binary merger events\;\cite{LIGOScientific:2016aoc,KAGRA:2021duu,KAGRA:2021vkt}, which can place strong constraints on PBH as dark matter in the mass range $\sim\mathcal{O}(1\textup{--}10^2){\rm\,M}_\odot$\;\cite{Bird:2016dcv,Sasaki:2016jop,Kovetz:2016kpi,Kovetz:2017rvv,Raidal:2017mfl,Takhistov:2017bpt,Ali-Haimoud:2017rtz,Bellomo:2017zsr,Raidal:2018bbj,Gow:2019pok,Hall:2020daa,Raidal:2024bmm,Andres-Carcasona:2024wqk,Bouhaddouti:2025ltb}. While constraints from PBH merger rates should be reexamined if PBHs were to form density spikes around SMBHs at galactic centers or clusters at high redshifts\;\cite{Feng:2024oab,DeLuca:2025nao,Aljaf:2025dta}.

Clustering of PBHs may evade stringent abundance constraints\;\cite{Carr:2016drx,Belotsky:2018wph,Carr:2020gox,DeLuca:2020jug,Franciolini:2022ewd,Stasenko:2024pzd} and produce distinctive gravitational‑wave signals\;\cite{Feng:2024oab,Stasenko:2025vqz,Zhang:2025tgm}. In addition, for the formation of SMBHs at high redshifts to happen from PBH runaway‑mergers, PBHs are required to be small-scale clustered\;\cite{Zhang:2025tgm}. The initial small-scale clustering of PBHs may arise from long–short mode coupling through various early-Universe mechanisms such as local-type non-Gaussianity\;\cite{Byrnes:2011ri,Byrnes:2012yx,Young:2013oia,Young:2014oea,Young:2015kda,Tada:2015noa,Franciolini:2018vbk,Desjacques:2018wuu,Ali-Haimoud:2018dau,Young:2019gfc,Suyama:2019cst,Atal:2020igj,DeLuca:2022bjs,DeLuca:2022uvz}, closed domain walls\;\cite{Khlopov:2004sc,Dokuchaev:2004kr,Belotsky:2018wph}, multistream inflation\;\cite{Ding:2019tjk,Huang:2023mwy}, long-range scalar forces\;\cite{Flores:2020drq}, quantum diffusion\;\cite{Ezquiaga:2019ftu,Ezquiaga:2022qpw,Animali:2024jiz}, or correlated bubble collisions\;\cite{DeLuca:2021mlh}.

The mechanism of small-scale PBH clustering offers several advantages: SMBHs can arise at sufficiently high redshifts—potentially explaining JWST observations of early galaxies and little red dots (LRDs)\;\cite{Matthee:2023utn,Pacucci:2023oci,Gentile:2024uiw,akins2024cosmos,williams2024galaxies}, as well as high‑redshift quasars\;\cite{Bogdan:2023ilu,Kovacs:2024zfh,Fu:2025aa}. Furthermore, this model circumvents late‑time limits on small‑mass PBHs and early‑Universe constraints on supermassive PBHs, i.e., CMB $\mu$‑distortion\;\cite{Carr:2016drx,Nakama:2017xvq,Carr:2020gox}, since low‑mass PBHs form early but the SMBH emerges via hierarchical mergers rather than direct collapse of primordial perturbations. Our analysis in Ref.\,\cite{Zhang:2025tgm} also shows that tidal interactions between PBH clusters can impart substantial angular momentum, enabling the formation of rapidly spinning central SMBHs and naturally producing extreme‑mass‑ratio inspirals (EMRIs). These events yield distinctive gravitational‑wave signatures that could be tested by future observatories\;\cite{Ruan:2018tsw,TianQin:2020hid,Kudoh:2005as,Kawamura:2020pcg,Colpi:2024xhw,ET:2025xjr} and distinguished from other direct collapse models\;\cite{Feng:2020kxv,Jiang:2025jtr,Shen:2025evo,Feng:2025rzf,Roberts:2025poo,Aggarwal:2025pit,Freese:2025dmo,Pacucci:2026ovn,Wang:2026vzg}.

To determine SMBH formation and cluster runaway timescales and to track mass‑distribution evolution, we employ \emph{the Smoluchowski coagulation equation} (``the Smoluchowski equation" hereafter for brevity) to model the dynamical evolution of a PBH cluster. The Smoluchowski equation, first formulated by Marian Smoluchowski in the early 20th century\;\cite{Smoluchowski:1916,Smoluchowski:1917}, is an integro‐differential equation that governs the time evolution of a size‐ or mass‐resolved particle distribution under binary coagulation. In the most general form, it reads\;\cite{Smoluchowski:1916,Smoluchowski:1917,Chandrasekhar:1943ws}
\begin{align}
\frac{\partial n(m,t)}{\partial t}
&=  \frac{1}{2}
  \int_{0}^{m}
    \mathcal{K}(m',\,m - m')\,n(m',t)\,n(m - m',t)\,\mathrm{d}m'
- n(m,t)
  \int_{0}^{\infty}
    \mathcal{K}(m,\,m')\,n(m',t)\,\mathrm{d}m'\,,
\end{align}
where \(n(m,t)\) is the number density of particles of mass \(m\) at time \(t\),  and \(\mathcal{K}(m,m')\) is the coagulation kernel characterizing the rate at which particles of masses \(m\) and \(m'\) adhere upon encounter. The first integral accounts for the gain in particles of mass \(m\) through coalescence of smaller pairs, while the second represents the loss due to further aggregation. Over the past century, this framework has been adapted to capture a wide range of phenomena—from the growth of colloidal aggregates in suspension\;\cite{russel1991colloidal,sauer1996theory}, the evolution of droplet size distributions in atmospheric aerosols\;\cite{Gillespie:1972,Gillespie:1975,agranovski2011aerosols} and the coalescence of planetesimals into planets\;\cite{Lee:2000,Malyshkin:2000ry} to hierarchical merging of dark matter halos\;\cite{Benson:2004gv,Neistein:2008ht,Erickcek:2006xc}. Its versatility lies in the choice of the kernel \(\mathcal{K}\), which can encode physical mechanisms such as shear‐driven collisions, gravitational focusing, or charge‐mediated sticking, making the Smoluchowski equation an indispensable tool across different fields.

The Smoluchowski equation is well suited to describe PBH cluster evolution. The merger of a black‑hole binary is driven by gravitational radiation following cluster virialization, so the kernel \(\mathcal{K}\) can be accurately modeled using the black‑hole merger cross section. In contrast to dark‑matter halo mergers, finite‑size effects of black holes are negligible, and a two‑body approximation remains valid at relevant cluster number densities, making the Smoluchowski equation particularly applicable. Moreover, compared to solving the modified Fokker–Planck equation\;\cite{Quinlan:1989}, the Smoluchowski equation features a simpler functional form. It effectively averages over the velocity and positional degrees of freedom of PBHs. When concerned solely with the mass distribution and runaway timescale—rather than detailed orbital dynamics and coalescence trajectories—the Smoluchowski equation substantially simplifies calculations while retaining accuracy. Furthermore, the Smoluchowski framework can be extended to accommodate more complex processes: cluster‑scale mass loss, black‑hole accretion during mergers, and the continuous formation of black holes throughout coalescence. Incorporating these extensions merits further investigation.

The Smoluchowski equation has been exploited to calculate black hole mergers within a star cluster, estimate the upper bound of the runaway timescale\;\cite{Mouri:2002mc}, and describe the PBH cluster dynamics\;\cite{Khlopov:2013zzo}. However, none of the studies performs detailed numerical calculations of the evolution of black hole mass distribution during the mergers. In this work, we present the first detailed simulation and application of the Smoluchowski equation to study the dynamical evolution runaway merger of PBH clusters. We numerically solve the equation by employing Monte Carlo simulation schemes. In our PBH cluster evolution problem, we adopt the full-conditioning scheme using the discrete inverse transform method\;\cite{Gillespie:1975,garcia_monte_1987,liffman_direct_1992}, which is more efficient compared to the partial-conditioning scheme using the acceptance-rejection method\;\cite{Gillespie:1975,wei_fast_2013,wei_gpubased_2013,xu_fast_2014,tran2023fragmentation}. We also develop optimized methods to significantly improve computational efficiency.

Our paper is organized as follows. In Section\;\ref{Sec:Smoluchowski}, we study the Smoluchowski equation, derive the PBH merger cross section, merger kernel, and discuss the mass segregation effect. In Section\;\ref{Sec:Monte_Carlo}, the full-conditioning scheme of Monte Carlo simulation for the Smoluchowski equation is provided. In Section\;\ref{Sec:simulations}, we present the simulation results, including the runaway timescale of a PBH cluster with and without mass segregation, and discuss the PBH mass distribution in a cluster at various redshifts, the evolution of the maximum PBH mass, as well as the merger rate. In Section\;\ref{Sec:discussion}, we discuss the implications and distinctive features of SMBH formation in the PBH clustering scenario. We summarize and conclude in Section\;\ref{Sec:conclusions}. Appendix\;\ref{App:proof_runaway} proves the upper limit of the runaway timescale. Appendix\;\ref{App:Sampling methods} provides the basic sampling methods for Monte Carlo simulations. Appendix\;\ref{App:partial-conditioning scheme} provides the basic of the partial-conditioning scheme in contrast to the full-conditioning scheme in Section\;\ref{Sec:Monte_Carlo}. Appendix\;\ref{App:optimization} details the optimization scheme implemented in our simulations. Appendix\;\ref{App:CPU_time} compares CPU times in partial- and full-conditioning schemes. 

\section{Smoluchowski coagulation equation}\label{Sec:Smoluchowski}
To determine the population distribution of a cluster of PBHs as a function of time, we consider solving the Smoluchowski \emph{coagulation} equation~\cite{Smoluchowski:1916,Smoluchowski:1917,Chandrasekhar:1943ws}:
\begin{equation}
\frac{{\rm d}}{{\rm d}t}n(m,t)=\frac{1}{2}\int_{m_{\rm min}}^m n(m',t)n(m-m',t)\mathcal{K}(m',m-m'){\rm\,d}m'
-n(m,t)\int_{m_{\rm min}}^{m_{\rm max}}n(m',t)\mathcal{K}(m,m'){\rm\,d}m',
\end{equation}
and set the upper and lower bounds of the integrals to be finite values, where $n(m,t){\rm\,d}m$ is the number density of PBHs in the mass range $(m, m+{\rm d}m)$ at time $t$, $\mathcal{K}(m_i,m_j)$ is the merger kernel of two PBHs $m_i$ and $m_j$, and $m_{\rm max/min}$ is the maximal/minimal BH mass during multiple consequential mergers. In the following the time dependence in $n$ is suppressed for convenience, and 
the coagulation equation in discretized form is written as:
\begin{equation}
\label{Eq:Smol_eq}
\boxed{
\frac{\rm d}{{\rm d}t}n_i=\frac{1}{2}\sum_{j+k=i}n_j n_k \mathcal{K}_{jk}
-n_i\sum_{j=1}^{N-i}n_j\mathcal{K}_{ij}.
}
\end{equation}
Here $n_i$ is the number density of PBH with mass $m_i=im_0~(i=1,2,...,N)$ and $m_0$ is the smallest PBH mass in the cluster. The first sum accounts for the increase of $n_i$ due to the merging of particles with $m_j+m_k=m_i$; while the second sum represents the decrease of $n_i$ due to the merging of particles of mass $m_i$ with those of any mass. For simplicity, we consider the initial mass distribution of black holes in the cluster is monochromatic.

\subsection{PBH merger cross section}
\label{Sec:GW_radiation}
Before solving the Smoluchowski equation for PBH clusters, we should first derive the merger cross section of two unbound black holes through the emission of gravitational waves. Assuming the initial separation of two black holes is much larger than their Schwarzschild radius in the cluster, we can treat black holes as point masses along the Newtonian trajectory.
If the two point masses, $m_i$ and $m_j$, far away with initial relative velocity $v_{\rm rel}$ pass close by, where the eccentricity $\texttt{e}\gtrsim1$, we can assume that it is close to a parabolic orbit and consider the leading term for $\texttt{e}\rightarrow1$.
The distance of closest approach of two massive objects is then given by
\begin{equation}
r_{\rm min}
=\frac{b^2v_{\rm rel}^2}{2G(m_i+m_j)},
\end{equation}
where 
$b$ is the impact parameter.
The quadrupole formalism gives the total energy of the gravitational radiation at Newtonian order
\begin{equation}
\delta E_{\rm GW}=\frac{8}{15}\frac{G^{7/2}}{c^5}\frac{(m_i+m_j)^{1/2}m_i^2m_j^2}{r_{\rm min}^{7/2}}g(\texttt{e})
\end{equation}
with $g(\texttt{e}=1)=425\pi/32\sqrt{2}$~\cite{Turner:1977}.
During the encounter, the condition for the two masses to form a gravitational-bound binary is that the energy loss must be larger than the kinetic energy in the center-of-mass frame, i.e.,
\begin{equation}
\delta E_{\rm GW}=\frac{85\pi}{12\sqrt{2}}\frac{G^{7/2}}{c^5}\frac{(m_i+m_j)^{1/2}m_i^2m_j^2}{r_{\rm min}^{7/2}}>
\frac{1}{2}\frac{m_im_j}{m_i+m_j} v_{\rm rel}^2.
\end{equation}
Thus one can solve for the maximum impact parameter $b_{\rm merg}$ for them to gravitationally bound, and obtain the merging cross section~\cite{Quinlan:1989,Mouri:2002mc} 
\begin{align}\label{Eq:cross_section}
\sigma_{\rm merg}\equiv\pi b_{\rm merg}^2
=2\pi\left(\frac{85\pi}{6\sqrt{2}}\right)^{2/7}\frac{G^2(m_i+m_j)^{10/7}m_i^{2/7}m_j^{2/7}}{c^{10/7}v_{\rm rel}^{18/7}}\;.
\end{align}
For a PBH cluster starting out with equal-mass PBHs, $m_1 = m_2 = m$, the merging cross section is 
\begin{align}
\sigma_{\rm merg}
\simeq
\left(\frac{85\pi}{3}\right)^{2/7}\sigma_{\rm scat}\left(\frac{v_{\rm rel}}{c}\right)^{10/7},
\end{align}
where $\sigma_{\rm scat}\simeq\pi\left(2Gm/v_{\rm rel}^2\right)^2$ is the gravitational (Coulomb) scattering cross section~\cite{Spitzer:1956hha,Spitzer:1987aa,Binney:1987gady}. It is straightforward to see that $\sigma_{\rm merg}\ll\left(c/v_{\rm rel}\right)^{10/7}\sigma_{\rm merg}\simeq\sigma_{\rm scat}$.
Therefore, a cluster of PBHs will result in only a tiny fraction of binary formation through multiple PBH scatterings during the relaxation process.

\subsection{The merger kernel and formulation}
The merger kernel $\mathcal{K}_{ij}$, governing the Smoluchowski equation Eq.\,\ref{Eq:Smol_eq}, is determined by the merging cross section of black holes $\sigma_{\rm merg}$ derived in Section\;\ref{Sec:GW_radiation} up to a relative velocity $v_{\rm rel}$. Because the equation is a mean-field approximation, the cross section has to be averaged over a velocity distribution. We assume that PBHs are in virial equilibrium through gravitational scattering and their \emph{relative} velocity follows a Maxwellian distribution, i.e.,
\begin{align}
P(v_{\rm rel},\alpha)=\sqrt{\frac{2\alpha^3}{\pi}}v_{\rm rel}^2\exp\left(-\frac{\alpha v_{\rm rel}^2}{2}\right)
\quad{\rm with}\quad\int_0^\infty P(v_{\rm rel},\alpha){\rm~d}v_{\rm rel}=1\,,
\end{align}
where the normalization coefficient $\alpha$ depends on the masses of two PBHs and the initial properties of the cluster. According to the equipartition of the kinetic energy among different mass populations during evolution, we have~\cite{Binney:1987gady,Mouri:2002mc}
\begin{align}
\frac{1}{2}m_iv_i^2=\frac{1}{2}m_0v_0^2,
\end{align}
where $v_i$ is the velocity dispersion of black holes with mass $m_i$, and $m_0$ and $v_0$ denote the reference mass and the corresponding velocity dispersion, 
which are the initial values of the cluster at formation. In addition, the kinetic energy of the two PBHs with relative velocity $v_{\rm rel}$ is given by
\begin{align}
E_{ij}=\frac{1}{2}\mu_{ij}v_{\rm rel}^2=\frac{1}{2}\frac{m_im_j}{m_i+m_j}v_{\rm rel}^2\,,
\end{align}
and the Boltzmann factor in the distribution is determined by
\begin{align}
\exp\left(-\frac{E_{ij}}{m_0v_0^2/3}\right)
=\exp\left(-\frac{\alpha v_{\rm rel}^2}{2}\right)
~{\rm with}~ \alpha\equiv\frac{3m_im_j}{(m_i+m_j)m_0v_0^2}\,,
\end{align}
where $m_0v_0^2/3$ is the initial virial ``temperature'' of the PBH cluster. Therefore, the PBH cluster evolves \emph{isothermally} during the scattering and sequential merger processes. 

The merger kernel is calculated as the velocity-averaged merging cross section with respect to their relative velocity
\begin{align}
\mathcal{K}_{ij}&=
\mathcal{K}(m_i,m_j)\equiv\langle\sigma_{\rm merg}(m_i,m_j)v_{\rm rel}\rangle
=\int_0^\infty P(v_{\rm rel}, \alpha)\,\sigma_{\rm merg}(m_i,m_j)\,v_{\rm rel}{\rm\,d}v_{\rm rel}\notag\\
&=\mathcal{A}\frac{G^2m_0^2}{c^3}\left(\frac{v_0}{c}\right)^{-11/7}\left(\frac{m_im_j}{m_0^2}\right)^{15/14}\left(\frac{m_i+m_j}{m_0}\right)^{9/14}\,,
\end{align}
where we have used Eq.~\ref{Eq:cross_section} for $\sigma_{\rm merg}(m_i,m_j)$ and $\mathcal{A}=85^{2/7}(2\pi)^{11/14}3^{1/2}\Gamma(5/7)$ with $\Gamma(z)\equiv\int_0^\infty t^{z-1}e^{-t}{\rm d}t$ the Gamma function. 
If the mass segregation is taken into account, the spatial distribution of PBHs must be specified when computing the merger rate. In this case, the merger rate will be modified by an additional factor $\mathcal{F}^{\rm ms}_{ij}$ according to the spatial distribution of PBHs, which will be discussed in the next subsection.

\paragraph{Numerical implementation}
When solving Eq.\,\ref{Eq:Smol_eq} numerically for PBHs of monochromatic mass $m_0$ with initial number density $n_0$ at $t=0$, we can rewrite the Smoluchowski equation in a dimensionless form by defining
\begin{equation}\label{Eq:dimensionless_quan_Smol}
\tilde{n}_i=n_i/n_0,
\quad
\tilde{\mathcal{K}}_{ij}=\mathcal{K}_{ij}/\mathcal{K}_{00}
\quad{\rm and}\quad
\tilde{t}=n_0\mathcal{K}_{00}t,
\end{equation}
where $\mathcal{K}_{00}=\mathcal{A}(G^2m_0^2/c^2)(v_0/c)^{-11/7}$. The mass of all PBHs produced by mergers is an integer multiple of the initial monochromatic PBH mass, denoted by $m_i = i m_0$ and thus $\tilde{\mathcal{K}}_{ij} = (ij)^{15/14}(i+j)^{9/14}$.
Then the evolution of number densities is given by
\begin{equation}\label{Eq:Smol_dimless}
\boxed{
\frac{\rm d}{{\rm d}\tilde{t}}\tilde{n}_i=\frac{1}{2}\sum_{j+k=i}\tilde{n}_j \tilde{n}_k \mathcal{\tilde{K}}_{jk}
-\tilde{n}_i\sum_{j=1}^{N-i}\tilde{n}_j\mathcal{\tilde{K}}_{ij}.
}
\end{equation}
As Eq.\,\ref{Eq:Smol_dimless} is dimensionless, all physical information: initial mass $m_0$, initial number density $n_0$, and initial velocity dispersion $v_0$, encoded in $n_0 \mathcal{K}_{00}$ will determine the timescale of the merger. This allows us to obtain the mass distribution of PBHs at any given time $\tilde{t}$ and the final merger time $\tilde{t}_{\text{final}}$. We then obtain the physical merger time by dividing the simulation time by $n_0 \mathcal{K}_{00}$.

\subsection{The mass segregation}
Since PBH scattering involves the exchange of energy and momentum, there is a statistical tendency for the kinetic energy of the two interacting PBHs to equalize during each encounter. This process ultimately leads to the equipartition of energy, which in turn drives mass segregation. In our model, we adopt a multimass approach, assuming that each mass population has its own velocity dispersion $v_i$. The effect of mass segregation becomes evident according to the equipartition,
\begin{align}\label{equipartition}
m_i v_i^2=m_j v_j^2\,,
\end{align}
meaning that populations with larger masses exhibit smaller velocity dispersions and tend to sink toward the center of the cluster through energy transfer. Assuming Boltzmann distribution for each population, this relation also allows us to derive the spatial distribution $f_i(r)$ for different mass components, i.e.,~\cite{Sigurdsson:1994ju,king1995mass}
\begin{align}
\label{Eq:f_mass_seg}
f_{i}(r)^{1/m_i}=f_{j}(r)^{1/m_j}\,,
\end{align} 
which directly demonstrates that when $m_i>m_j$, the distribution of heavier components is more concentrated at the center, while lighter components are more spread out from the center.

As we have mentioned, the Smoluchowski equation is a mean-field approximation that contains only temporal information. The densities of different mass components in a cluster are averaged over velocity and spatial fluctuations, with all the physical information encoded in the kernel $\mathcal{K}_{ij}$. In the previous discussion, we take velocity-averaged of the mergining cross section using a Maxwellian distribution but do not account for the effects of mass segregation, which can lead to a spatial distribution of components that evolve over time. To preserve the structure of the equation and capture the dynamics of mass segregation, we need to account for the spatial average in the cluster. 

On the right-hand side of Eq.\,\ref{Eq:Smol_eq}, the merger effect consists of two parts: the first part represents the increase in PBH number density, and the second part describes the decrease. Here, we focus on the first term, while the second term follows similarly. Specifically, we compute
\begin{align}
\int n_i(t, r)n_j(t, r)\langle\sigma_{\text{merg}}v_{\text{rel}}\rangle {\rm\,d}^3\mathbf{r} 
=4\pi\int  n_i(t, r)n_j(t, r)\,r^2{\rm\,d}r\, \langle\sigma_{\text{merg}}v_{\text{rel}}\rangle
\end{align}
where $n_i(t, r)$ is the number density of PBHs with mass $m_i$ at $r$ given time $t$ and the angle brackets indicates averaging over $v_{\text{rel}}$ as before. We assume that the cluster is isotropic, so the integration is performed only over $r$, and $\langle\sigma_{\text{merg}}v_{\text{rel}}\rangle$ factors out of the spatial integral as it is independent of the spatial distribution. Besides, we assume that the number density factorizes as $n_i(t,r)=\hat{n}_{i}(t)f_{i}(r)$ so that the temporal dependence is contained in $\hat{n}_{i}$, while the spatial profile $f_i$ remains time-independent. 
We can then extract the mean number densities of $i$ and $j$ populations, leaving the remaining part as the spatially averaged effective kernel, which incorporates the effects of mass segregation, i.e.,
\begin{align}
4\pi&\int  n_i(t, r)n_j(t, r)\,r^2{\rm\,d}r\equiv 
V\langle n_i(t,r)n_j(t,r)\rangle_r
=n_i(t)n_j(t)\frac{V\langle f_i(r)f_j(r)\rangle_r}{\langle f_i(r)\rangle_r \langle f_j(r)\rangle_r}\,,
\end{align}
where $V=\int{\rm d}^3\mathbf{r}$ is the volume of a cluster and the subscript ``$r$'' indicates the spatial average, and we denote $n_i(t)\equiv\hat{n}_i(t)\langle f_{i}(r)\rangle_{r}$ as the new averaged variable. Thus, the factor 
\begin{equation}
\mathcal{F}^{\rm ms}_{ij}=\frac{V\langle f_i(r)f_j(r)\rangle_r}{\langle f_i(r)\rangle_r \langle f_j(r)\rangle_r}
\end{equation} 
is attached to the merger kernel $\mathcal{K}_{ij}=\langle\sigma_{\rm merg}(m_i,m_j)v_{\rm rel}\rangle$ without mass segregation, such that
\begin{equation}
\mathcal{K}_{ij}=\langle\sigma_{\rm merg}(m_i,m_j)v_{\rm rel}\rangle\mathcal{F}^{\rm ms}_{ij}
\end{equation}
is a modification of the effect of spatial distribution.
To obtain an analytical spatial distribution for $m_i$, the initial distribution of PBHs for $m_0$ in a cluster should be specified. We further assume that the distribution for each mass population remains approximately the same according to Eq.\,\ref{Eq:f_mass_seg} throughout the cluster evolution. In our semi-analytical study, we consider Gaussian and Plummer distributions as the initial PBH distribution to model the effect of mass segregation.

\paragraph{Gaussian distribution}
The Gaussian profile is commonly used to fit observed surface-brightness profiles of open clusters or galactic nuclei when the outer regions are unresolved\;\cite{Peng:2009aa, Schodel:2014aa}. The density profile reads:
\begin{align}
f_0=\mathrm{exp}\left(-\frac{r^2}{2r_0^2}\right),
\end{align}
where $r_0$ is the scale of the cluster. The spatial distribution of $m_i$ according to Eq.\,\ref{Eq:f_mass_seg} yields:
\begin{align}
f_i=f_0^{m_i/m_0}=\mathrm{exp}\left(-\frac{m_i}{m_0}\frac{r^2}{2r_0^2}\right),
\end{align}
and we obtain the enhancement factor induced by mass segregation
\begin{align}
\mathcal{F}^{\rm ms}_{ij}=\frac{V\langle f_i(r)f_j(r)\rangle_r}{\langle f_i(r)\rangle_r \langle f_j(r)\rangle_r} =2\sqrt{2}\left(\frac{m_im_j}{m_0\left(m_i+m_j\right)}\right)^{3/2}\;.
\end{align}
It is worth noting that, here \(V\) is treated as a free parameter with the dimensions of volume. In this work we choose \(V\) so that when \(m_i = m_j = m_0\) the prefactor equals unity, i.e. the expression reduces to the case without mass segregation. In reality \(V\) should be calibrated against \(N\)-body simulations; we postpone such a calibration to future work. The same prescription is adopted in the Plummer-case calculation below.

\paragraph{Plummer distribution} 
On the other hand, the Plummer model, which was first used to fit the observation of globular clusters and is widely used to describe spherical star clusters and galaxies\;\cite{Plummer:1911zza,dejonghe1987completely}. 
The density profile reads:
\begin{align}
f_0=\left(1+\frac{r^2}{r_0^2}\right)^{-5/2}\;.
\end{align}
The spatial distribution of $m_i$ according to Eq.\,\ref{Eq:f_mass_seg} is:
\begin{align}
f_i=f_0^{m_i/m_0}=\left(1+\frac{r^2}{r_0^2}\right)^{-5m_i/2m_0}\,,
\end{align}
and the mass-segregation factor becomes 
\begin{align}
\mathcal{F}^{\rm ms}_{ij}=\frac{V\langle f_i(r)f_j(r)\rangle_r}{\langle f_i(r)\rangle_r \langle f_j(r)\rangle_r} =\left(\frac{256}{15\pi}\right)\frac{\Phi\left[\frac{5}{2}\frac{m_i+m_j}{m_0}\right]}{\Phi\left[\frac{5}{2}\frac{m_i}{m_0}\right]\Phi\left[\frac{5}{2}\frac{m_j}{m_0}\right]}\,,
\end{align}
where
\begin{align}
\Phi[\beta]=\frac{\Gamma\left[5/2\right]\Gamma\left[\beta-3/2\right]}{\Gamma\left[\beta\right]}\;.
\end{align}

When comparing the two profiles, the Plummer model produces a larger enhancement factor as the black hole mass increases. This implies that, under the Plummer profile, mergers of more massive black holes occur more rapidly, thereby accelerating the overall merger process and reducing the runaway timescale, which will be demonstrated in Section\;\ref{Sec:simulations}. 
In addition, it has been shown in Ref.\;\cite{Mouri:2002mc} that, in the absence of mass segregation, there exists an upper bound on the runaway timescale. We show that the corresponding upper bound also exists in the case with mass segregation. More generally, if the merger kernel \(\mathcal{K}_{ij}\) admits a lower bound given by a convex, positively homogeneous function of degree \(\lambda\), i.e.,
$\mathcal{K}(s i, s j)=s^{\lambda}\mathcal{K}(i,j)$ for $s>0$, then \(\lambda>1\) implies the existence of a finite runaway timescale. The proof can be carried out rigorously in the \(N\to\infty\) limit (see Appendix\;\ref{App:proof_runaway}). For finite-size systems, the numerical simulations in Section~\ref{Sec:simulations} confirm that the runaway timescales are finite and cases including mass segregation exhibit shorter runaway timescales than that without mass segregation.

\section{Monte Carlo simulation}
\label{Sec:Monte_Carlo}
The discrete Smoluchowski equation is a set of coupled nonlinear ordinary differential equations. Depending on the initial number of PBHs in the cluster, the number of equations can reach $10^4$ or more, making direct solving difficult and resource-intensive. 

In this paper, we adopt the Monte Carlo numerical method to solve the Smoluchowski equation, as it is well-suited for inherently discrete processes and does not require prior knowledge of the PBH mass distribution. Using this method, we can simulate a stochastic process and obtain the PBH mass distribution of a cluster at any given merger time.

In the following, we review the concepts and numerical method in Refs.\,\cite{Gillespie:1972,Gillespie:1975,garcia_monte_1987}.
For a PBH cluster with an initial number density $n_0\equiv N/V$, the merger kernel is given by
\begin{align}
\frac{\mathcal{K}(\texttt{m}_i, \texttt{m}_j)\,n_0}{N} \, {\rm d}t = \frac{\mathcal{K}(\texttt{m}_i, \texttt{m}_j)}{V} \, {\rm d}t \equiv \frac{\mathcal{K}_{ij}}{V} \, {\rm d}t\,,
\end{align}
which represents the probability that a given pair of PBHs with masses \( \texttt{m}_i \) and \( \texttt{m}_j \) will merge in the next infinitesimal time interval \( {\rm d}t \). Note that here \( i \) denotes the PBH at the \(i\)-th element in a mass array, \([\texttt{m}_1, \texttt{m}_2, \dots, \texttt{m}_{N}]\), constructed for simulations, and \(\texttt{m}_i\) is the corresponding PBH mass, which differs from that of \( m_i = i m_0 \) defined in Section\;\ref{Sec:Smoluchowski}. We assume that the merger timescales of PBH binaries are sufficiently short to neglect perturbations from other PBHs during the merger process. Consequently, the merger rate can be approximated by the PBH binary formation rate, with the merger process treated as instantaneous. Initially, we consider \( N \) PBHs, labeled \( i \). First, two PBHs are selected through sampling, and when they can merge, the mass of one is set to zero, and the other is updated as the sum of the two masses. Then the mass distribution is updated, and the above steps are repeated until only one black hole remains.

\subsection{The merger probability density function}
\begin{figure}[t!]
\centering	\includegraphics[width=0.6\textwidth]{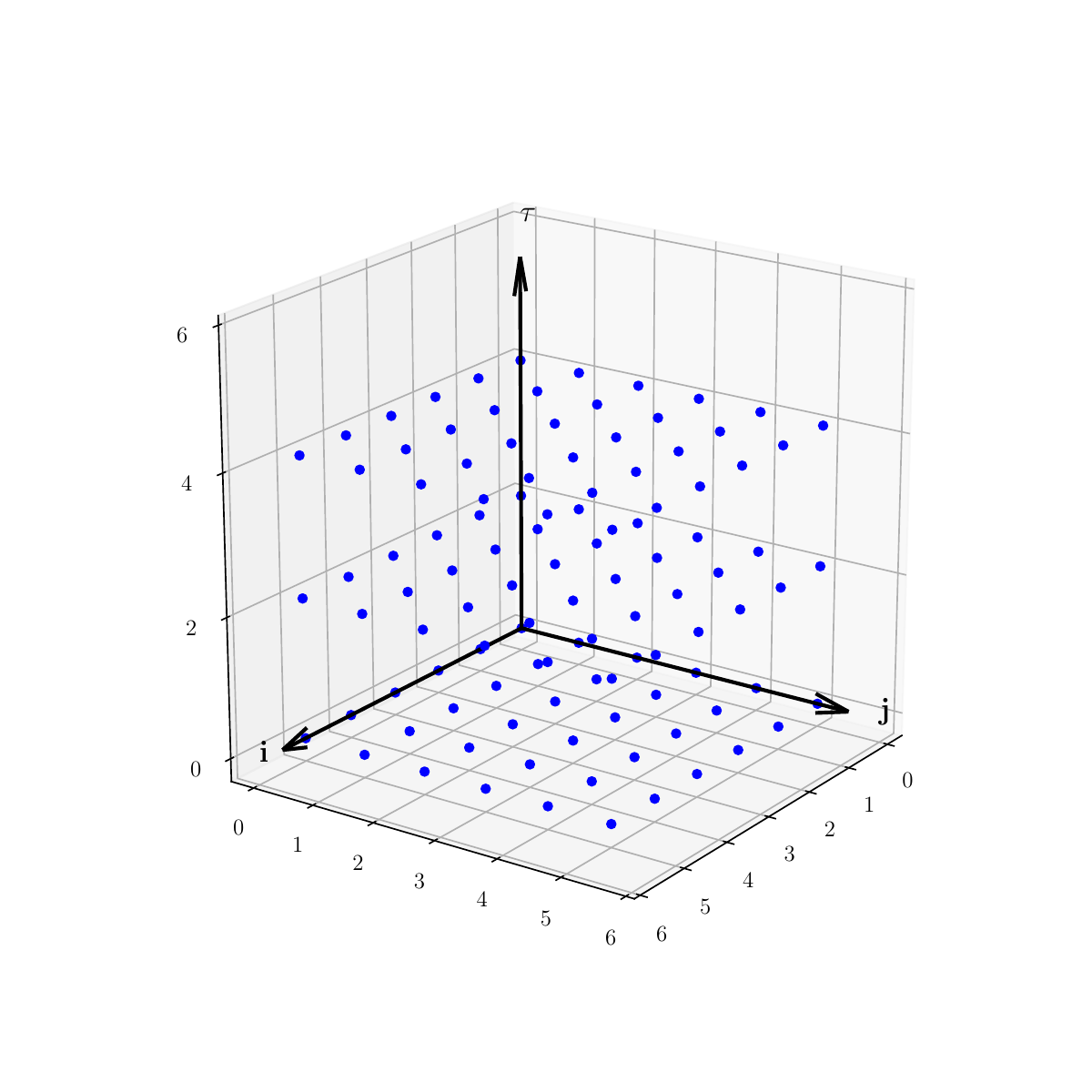}
\caption{Schematic plot of the domain of the function $P(\tau,i,j)$}\label{Fig:domain_P}
\end{figure}

To perform Monte Carlo sampling, we define the merger probability density function\;\cite{Gillespie:1972,Gillespie:1975,garcia_monte_1987}:
\begin{align}
P(\tau, i, j) {\rm\,d}\tau = \text{the~probability~at~time~}t\notag
\end{align}
\begin{align}
\text{~that~the~next~merger~will~occur~in~the~time interval~}
\end{align}
\begin{align}
[t + \tau, t + \tau + {\rm d}\tau] \text{~and~will~involve~PBHs~} i \text{ and } j\;.\notag
\end{align}
This function is calculated as the product of three probabilities: the probability $P_a$ that none of the PBHs merge in the interval $[t, t + \tau]$, the probability $P_b$ that PBHs $i$ and $j$ merge in the next interval $[t + \tau, t + \tau + {\rm d}\tau]$, and the probability $P_c$ that no other PBHs merge in the same infinitesimal time interval ${\rm d}\tau$. The domain of the function $P(\tau,i,j)$ is shown in Fig.\,\ref{Fig:domain_P}.

To derive the first probability $P_a$, we divide the time interval $[t, t + \tau]$ into $q \in \mathbb{N}$ subintervals $\delta = \tau/q$. When $q$ is sufficiently large, the probability that $k$ and $l$ will not merge in the first $\delta$-interval is $(1 - \mathcal{K}_{kl} \delta/V)$. Thus, the probability that all PBH pairs will not merge in any of the $q$ $\delta$-intervals in $[t, t + \tau]$ is approximately:
\begin{align}
\prod_{k=1}^{N-1} \prod_{l=k+1}^{N} \left(1 - \frac{\mathcal{K}_{kl}}{V} \delta \right)^q.
\end{align}
Taking the limit $q \to \infty$, we have:
\begin{align}
\lim_{q \to \infty} \prod_{k=1}^{N-1} \prod_{l=k+1}^{N} \left(1 - \frac{\mathcal{K}_{kl}}{V} \delta \right)^q = \prod_{k=1}^{N-1} \prod_{l=k+1}^{N} \exp\left( -\frac{\mathcal{K}_{kl}}{V} \tau \right) = \exp\left( -\sum_{k=1}^{N-1} \sum_{l=k+1}^{N} \frac{\mathcal{K}_{kl}}{V} \tau \right).
\end{align}
Thus, the probability $P_a$ becomes:
\begin{align}
P_a = \exp\left( -\sum_{k=1}^{N-1} \sum_{l=k+1}^{N} \frac{\mathcal{K}_{kl}}{V} \tau \right).
\end{align}
Next, the probability $P_b$ that PBHs $i$ and $j$ merge in the next interval $[t + \tau, t + \tau + {\rm d}\tau]$ is given by:
\begin{align}
P_b = \frac{\mathcal{K}_{ij}}{V}{\rm\,d}\tau.
\end{align}
The probability $P_c$ that no other PBH pairs merge in the same infinitesimal time interval ${\rm d}\tau$ is:
\begin{align}
P_c = \prod_{\substack{k=1 \\ k \neq i}}^{N-1} \prod_{\substack{l=k+1 \\ l \neq j}}^{N} \left( 1 - \frac{\mathcal{K}_{kl}}{V} {\rm\,d}\tau \right) \approx \exp\left( - \sum_{\substack{k=1 \\ k \neq i}}^{N-1} \sum_{\substack{l=k+1 \\ l \neq j}}^{N} \frac{\mathcal{K}_{kl}}{V} {\rm\,d}\tau \right) \approx 1.
\end{align}
Thus, the joint probability density function $P(\tau, i, j)$ is given by:
\begin{align}\label{Ptauij}
P(\tau, i, j) = \frac{\mathcal{K}_{ij}}{V} \exp\left( - \sum_{k=1}^{N-1} \sum_{l=k+1}^{N} \frac{\mathcal{K}_{kl}}{V} \tau \right), \quad 0 \leq \tau < \infty, \quad 1 \leq i < j \leq N.
\end{align}
Having obtained this probability density function, we can use it to sample the time interval $\tau$ corresponding to each merge event and the pair of PBHs $i$ and $j$. The choice of sampling scheme is crucial for our simulation efficiency. Two basic approaches are commonly used: the partial-conditioning scheme implemented with the \emph{acceptance-rejection method}\;\cite{Gillespie:1975,wei_fast_2013,wei_gpubased_2013,xu_fast_2014,tran2023fragmentation}, and the full-conditioning scheme implemented with the \emph{discrete inverse transformation method}\;\cite{Gillespie:1975,garcia_monte_1987,liffman_direct_1992}. For the hierarchical merger-rate structure that arises in our problem, the latter is more appropriate (see Appendix\;\ref{App:partial-conditioning scheme} for a detailed discussion). We therefore focus on the full-conditioning scheme in the following. We introduce the method, describe its implementation for Monte Carlo simulations of PBH-cluster mergers, and explain its advantages. The prescription of the partial-conditioning approach is provided in Appendix\;\ref{App:partial-conditioning scheme} for completeness.

\subsection{Full-conditioning Monte Carlo scheme}
In this section, we adopt the full-conditioning scheme using the discrete inverse transformation method to perform the simulation. This method is suitable and exact for any probability distribution with a discrete set of states. Since this method usually requires calculating the cumulative sum, it is not efficient when the number of components is large. However, in our model, due to the huge hierarchy of merger rate during the merger, this method can be more efficient than the partial-conditioning scheme using the acceptance-rejection method, e.g. when $N\gtrsim \mathcal{O}(10^4)$. In our work, the computational efficiency can be effectively improved by the ``unique-mass-mapping method'' with the book-keeping technique (see Appendix\;\ref{App:optimization}). 

The ``full-conditioning'' scheme means that the three-variable probability density function \(P(\tau,i,j)\) is decomposed using the chain rule of conditional probability:
\begin{align}
P(\tau, i, j) = P_1(\tau) \cdot P_2(i|\tau) \cdot P_3(j|\tau, i)
\end{align}
so that the joint distribution is expressed as a product of three single-variable (possibly conditional) factors, where 
$P_1(\tau){\rm\,d}\tau$ is the probability that the next merger will happen in the time interval $[t + \tau, t + \tau + {\rm d}\tau]$, independent of which pair of PBHs is chosen.
 $P_2(i|\tau)$ is the probability that the next merger will include the PBH $i$, given that it occurs at $t + \tau$. The final probability $P_3(j|\tau, i)$ means that, given that the merger occurs at $t + \tau$ and PBH $i$ is involved, the probability that PBH $j$ is also involved. These probabilities are given by:
\begin{align}\label{Eq:prob_full}
P_1(\tau) = \frac{\mathcal{K}_0}{V} \exp \left( -\frac{\mathcal{K}_0}{V} \tau \right)\,, \quad P_2(i|\tau) = P_2(i) = \frac{\mathcal{K}_i}{\mathcal{K}_0}\,, \quad P_3(j|i, \tau) = P_3(j|i) = \frac{\mathcal{K}_{ij}}{\mathcal{K}_i}\,,
\end{align}
where the {cumulative} sums $\mathcal{K}_i$ and the normalization factor $\mathcal{K}_0$ are
\begin{align}
\mathcal{K}_i = \sum_{j=i+1}^{N} \mathcal{K}_{ij}, \quad \mathcal{K}_0 = \sum_{i=1}^{N-1} \mathcal{K}_i = \sum_{i=1}^{N-1} \sum_{j=i+1}^{N} \mathcal{K}_{ij}
\end{align}
such that the probabilities defined above are all normalized
\begin{align}
\int_0^\infty P_1(\tau){\rm\,d}\tau=1\,, \quad \sum_{i=1}^{N-1} P_2(i|\tau)=1\,,~{\rm and} \quad \sum_{j=i+1}^{N}P_3(j|i, \tau)=1
\end{align}
and the joint probability density is also normalized
\begin{align}
\int_0^\infty {\rm d}\tau \sum_{i=1}^{N-1} \sum_{j=i+1}^{N} P(\tau, i, j)=1\;.
\end{align}
After separating the probability density, we can perform the sampling for $\tau$, $i$, and $j$. We can generate $\tau$ using the \emph{inverse transform method} (see Appendix\;\ref{Append:inversion transformation method}):
\begin{align}
\tau = \frac{V}{\mathcal{K}_0} \ln \left( \frac{1}{r_1} \right),
\end{align}
where $r_1 \in [0, 1]$ is a uniformly distributed random number.
Then, we generate a pair of PBHs using the \emph{discrete inverse transformation method} given the cumulative sums $\mathcal{K}_i$ and $\mathcal{K}_0$. Specifically, a random integer $i$ is generated from $P_2(i)$ with $r_2 \in [0, 1]$, and after obtaining $i$, a random integer $j$ can be generated from $P_3(j|i)$ with $r_3 \in [0, 1]$:
\begin{align}
\sum_{i'=1}^{i-1} \mathcal{K}_{i'} < r_2 \mathcal{K}_0 \leq \sum_{i'=1}^{i} \mathcal{K}_{i'}, \quad \sum_{j'=i+1}^{j-1} \mathcal{K}_{ij'} < r_3 \mathcal{K}_i \leq \sum_{j'=i+1}^{j} \mathcal{K}_{ij'}.
\end{align}

In the actual simulation we solve the dimensionless Smoluchowski equation, in which the dimensionless quantities are defined as in Eq.\,\ref{Eq:dimensionless_quan_Smol}: \(\tilde{n}_i=n_i/n_0,\;\tilde{\mathcal{K}}_{ij}=\mathcal{K}_{ij}/\mathcal{K}_{00}\;{\rm and}\;\tilde{t}=n_0\mathcal{K}_{00}t\).

\subsubsection*{The basic simulation algorithm}
After completing the sampling process, the simulation algorithm proceeds as follows:
\begin{itemize}

\item Step 0: Initialize the time to \(\tilde{t}_0 = 0\) and  the number density \(\tilde{n}_0 = 1\) by construction. Specify masses for \(N\) PBHs as 
    \begin{align}
      \mathbf{mass} = [\texttt{m}_1, \texttt{m}_2, \dots, \texttt{m}_N].
    \end{align}
    For a monochromatic distribution, \(\mathbf{mass} = [m_0, m_0, \dots, m_0]\); in the dimensionless unit, this becomes \(\mathbf{mass} = [1, 1, \dots, 1]\) (absent PBHs are denoted by \(\texttt{m}_i = 0\)). Then, compute the kernels \(\tilde{\mathcal{K}}_i\) and \(\tilde{\mathcal{K}}_0\). Then, construct the unique‑mass array—initially \([1]\) for the monochromatic case—and establish the mapping from the full mass array to this unique-mass array.

\item Step 1: According to the probability density $P(\tilde{\tau}, i, j)$, generate a random triplet $(\tilde{\tau}, i, j)$ by employing the full-conditioning scheme with the discrete inverse transformation method.

\item Step 2: For the selected PBH pair \((i,j)\), merge masses by setting \(\texttt{m}_i + \texttt{m}_j\rightarrow\texttt{m}_{j}\) and set \(\texttt{m}_i=0\). Next, advance the dimensionless time \(\tilde{t}\) by \( \tilde{\tau} = \left(N/\tilde{n}_0\,\tilde{\mathcal{K}}_0\right)\ln\left(1/r_1\right)\). Then update the unique‑mass array, the mass‑mapping matrix, \(\tilde{\mathcal{K}}_i\) and \(\tilde{\mathcal{K}}_0\) using the bookkeeping routine.

\item Step 3:  Stop the calculation if the total number of PBHs is $1$ or if $\tilde{t} \geq \tilde{t}_f$ with \(\tilde{t}_f\) being the time when the simulation is forced to end. Otherwise, return to Step 1.
\end{itemize}

\section{Simulation results}
\label{Sec:simulations}
We study the evolution of PBH clusters over a range of cluster sizes, \(N=N_{\rm cl} = 10^2\)–\(10^6\), considering both cases with and without mass segregation, and evaluate the corresponding runaway timescales. For cases with mass segregation, we consider both Gaussian and Plummer models.

\subsection{The runaway timescale}
The runaway timescale characterizes the formation of SMBHs during cluster evolution. In the Monte Carlo simulations, the merger rate first increases when the runaway merger happens, and the entire cluster eventually merges into a single object. However, when the number of PBHs becomes sufficiently small, the Monte Carlo simulation—and, a fortiori, the Smoluchowski equation—ceases to be valid, because the equation is a statistical description that requires a sufficiently large sample. Therefore, we terminate the simulation when the maximal PBH reaches half of the total cluster mass, i.e., 
\begin{align}
m_{\rm cut}\simeq \frac{1}{2}M_{\rm cl}\,,
\end{align}
which defines the runaway timescale. The subsequent evolution of the cluster should be complemented by $N$-body simulations which we leave for the future work.

\begin{figure}[htbp]
\centering
\includegraphics[width=0.8\textwidth]{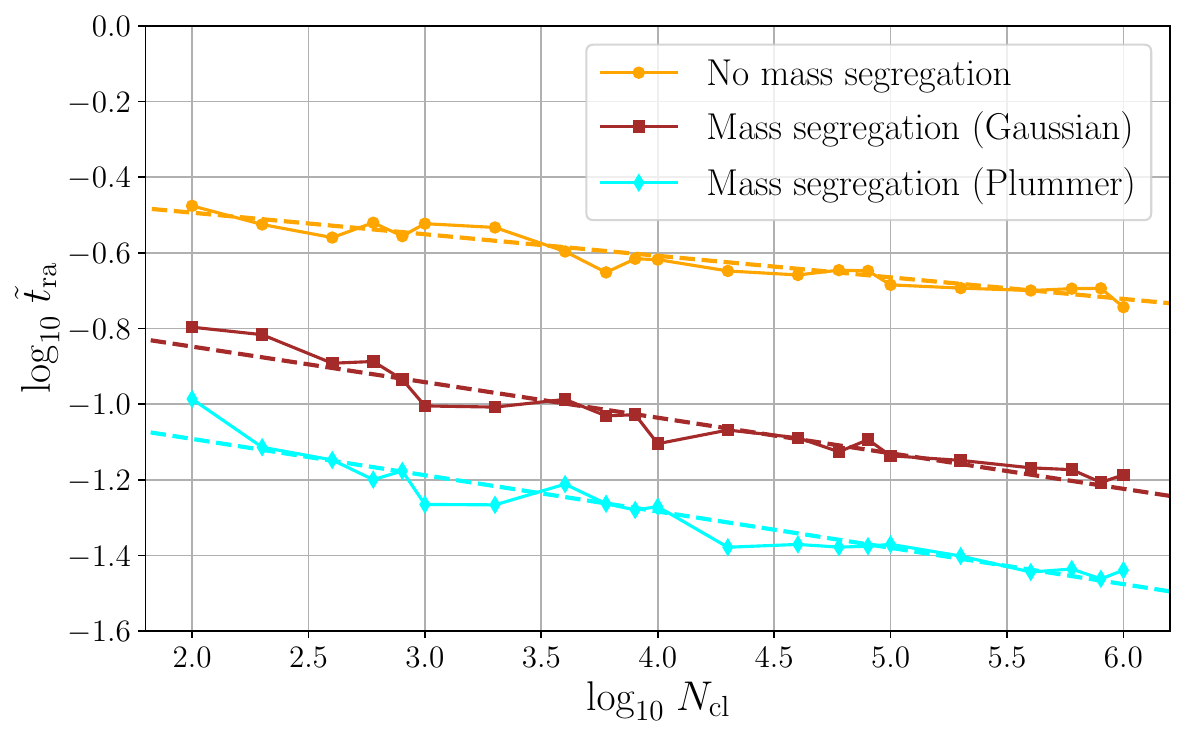}
\caption{Dimensionless runaway timescales as a function of \(N_{\rm cl}\). Orange circles show results without mass segregation, brown squares correspond to the Gaussian model, and cyan diamonds correspond to the Plummer model. The corresponding dashed lines are linear fits for each model. The dimensionless runaway timescale is defined as the time at which the central SMBH attains approximately half of the total cluster mass.}
\label{Fig:runaway_timescale}
\end{figure}

Fig.\,\ref{Fig:runaway_timescale} demonstrates the runaway timescales $\tilde{t}_{\rm ra}$ with and without mass segregation as a function of the number of PBHs $N_{\rm cl}$ in a cluster. Generally, $\tilde{t}_{\rm ra}$ decreases with $N_{\rm cl}$ due to the enhanced merger rate. While with mass segregation, $\tilde{t}_{\rm ra}$ is reduced by a factor of $\sim 2$--$5$. The relationship between $\tilde{t}_{\rm ra}$ and $N_{\rm cl}$ is approximately linear in the $\log$ space. To quantify the scaling in the log space, we perform the fit for each case:
\begin{align}
  & \text{No mass segregation}:~ \log_{10}\left(\tilde{t}_{\rm ra}\right)=-0.38-0.057\log_{10}N_{\rm cl}\;.\\
  \notag\\
  & \text{Mass segregation (Gaussian)}:~ \log_{10}\left(\tilde{t}_{\rm ra}\right)=-0.66-0.094\log_{10}N_{\rm cl}\;.\\
  \notag\\
  & \text{Mass segregation (Plummer)}:~ \log_{10}\left(\tilde{t}_{\rm ra}\right)=-0.90-0.096\log_{10}N_{\rm cl}\;.
\end{align}
The runaway timescale with mass segregation is reduced because more massive black holes tend to transfer energy to less massive ones and sink toward the cluster center, whereas the less massive black holes gain energy and move outward. Since the merger cross section increases with mass, the timescale for forming a SMBH through mergers is consequently shortened. In addition, compared to the Gaussian model, the Plummer model results in a shorter runaway timescale.

Furthermore, Fig.\,\ref{Fig:runaway_timescale} also shows that the runaway timescales obtained from the simulations lie below the analytical upper bound derived in Ref.\,\cite{Mouri:2002mc} (see also Appendix\;\ref{App:proof_runaway}), demonstrating consistency between our numerical results and the mathematical analysis of the Smoluchowski equation. 
As discussed in Section\;\ref{Sec:Smoluchowski}, we multiply the dimensionless $\tilde{t}_{\rm ra}$ with $n_0\mathcal{K}_{00}$ to obtain the physical runaway timescale.

\subsection{The PBH mass distribution at different cosmic redshifts} 
By Monte Carlo simulations, we can obtain the cluster evolution history, including the PBH mass distribution, the growth of the maximal black hole, and the total merger rate. This is crucial for understanding the structural evolution of PBH clusters and the associated gravitational-wave signals\;\cite{Zhang:2025tgm}. 

As in the physical setup of Ref.\,\cite{Zhang:2025tgm}, we consider that the PBH clusters form in the radiation-dominated era and SMBHs emerge through runaway mergers in the matter-dominated era. The cosmic time is computed as
\begin{equation}
t(z)=\frac{2}{3 H_0 \sqrt{\Omega_\Lambda}} \ln\left[ \frac{1}{(1 + z)^{3/2}}\left( \frac{\Omega_\Lambda}{\Omega_{\rm M}} \right)^{1/2} + \sqrt{1 + \frac{1}{(1 + z)^3}\frac{\Omega_\Lambda}{ \Omega_{\rm M}}} \right]\,,
\end{equation}
where $\Omega_\Lambda\simeq0.688$, $\Omega_{\rm M}\simeq0.312$ and $H_0\simeq67.6{\rm\,km\,s^{-1}\,Mpc^{-1}}$ from Planck cosmology\;\cite{Planck:2018vyg}. 
In the following, we relabel $n_0=n_{\rm cl}$ and $m_0=m_{\rm pbh}$ for PBH clusters specifically.

\begin{figure}[htbp]
    \centering
    \begin{minipage}{0.49\linewidth}
    \centering
    \includegraphics[width=0.9\linewidth]{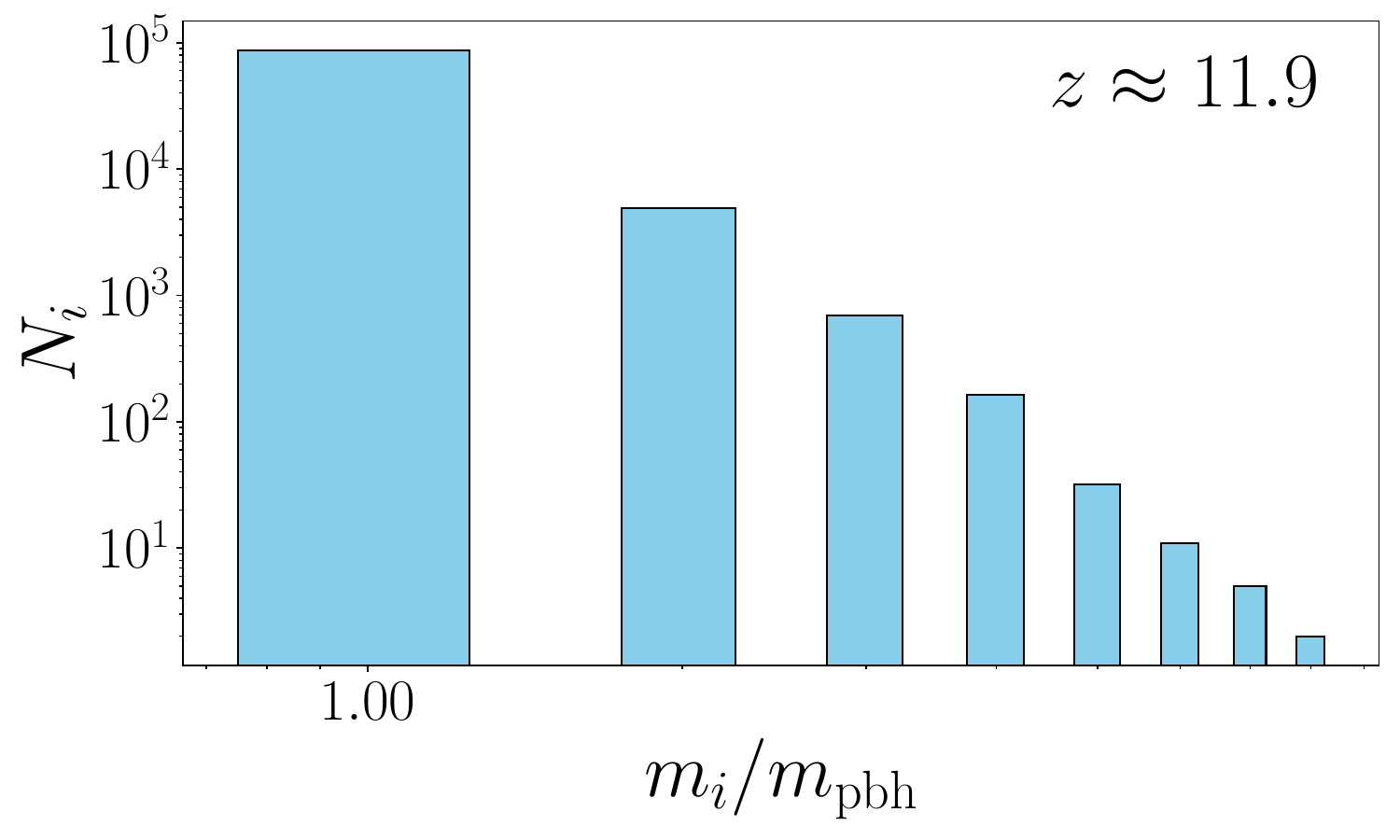}
    \label{mass_dis_1} 
    \end{minipage}
    \begin{minipage}{0.49\linewidth}
    \centering
    \includegraphics[width=0.9\linewidth]{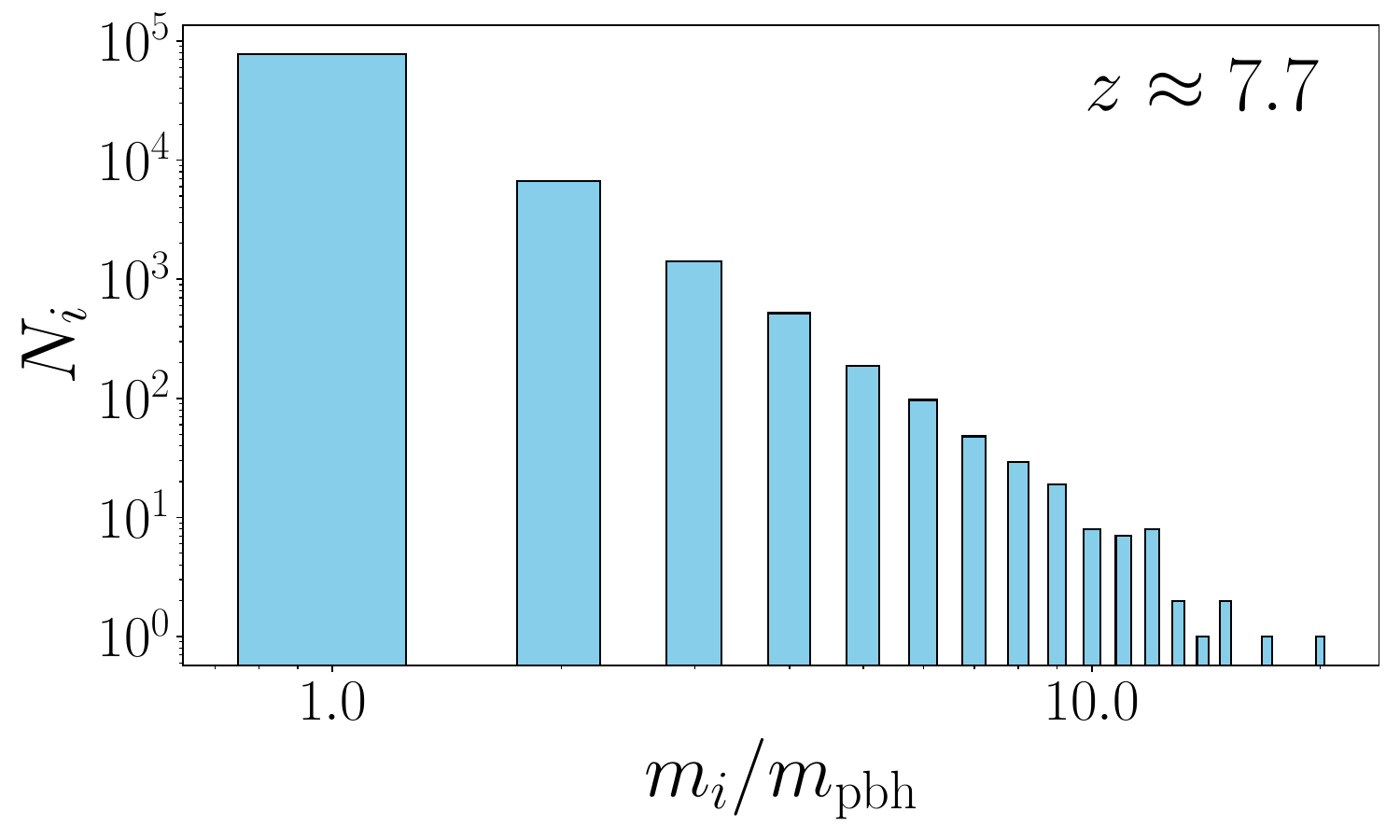}
    \label{mass_dis_2} 
    \end{minipage}
    \begin{minipage}{0.49\linewidth}
    \centering
    \includegraphics[width=0.9\linewidth]{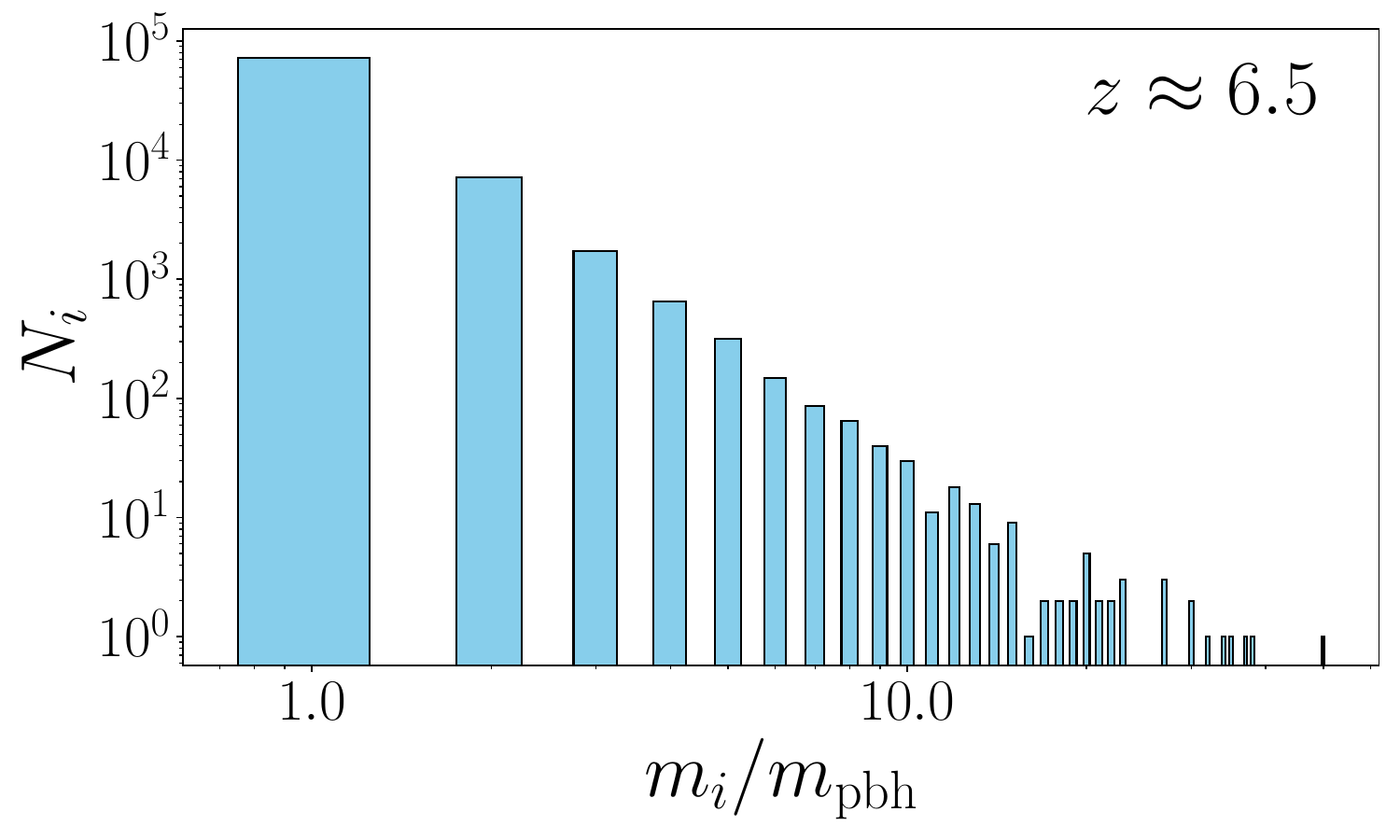}
    \label{mass_dis_3} 
    \end{minipage}
    \begin{minipage}{0.49\linewidth}
    \centering
    \includegraphics[width=0.9\linewidth]{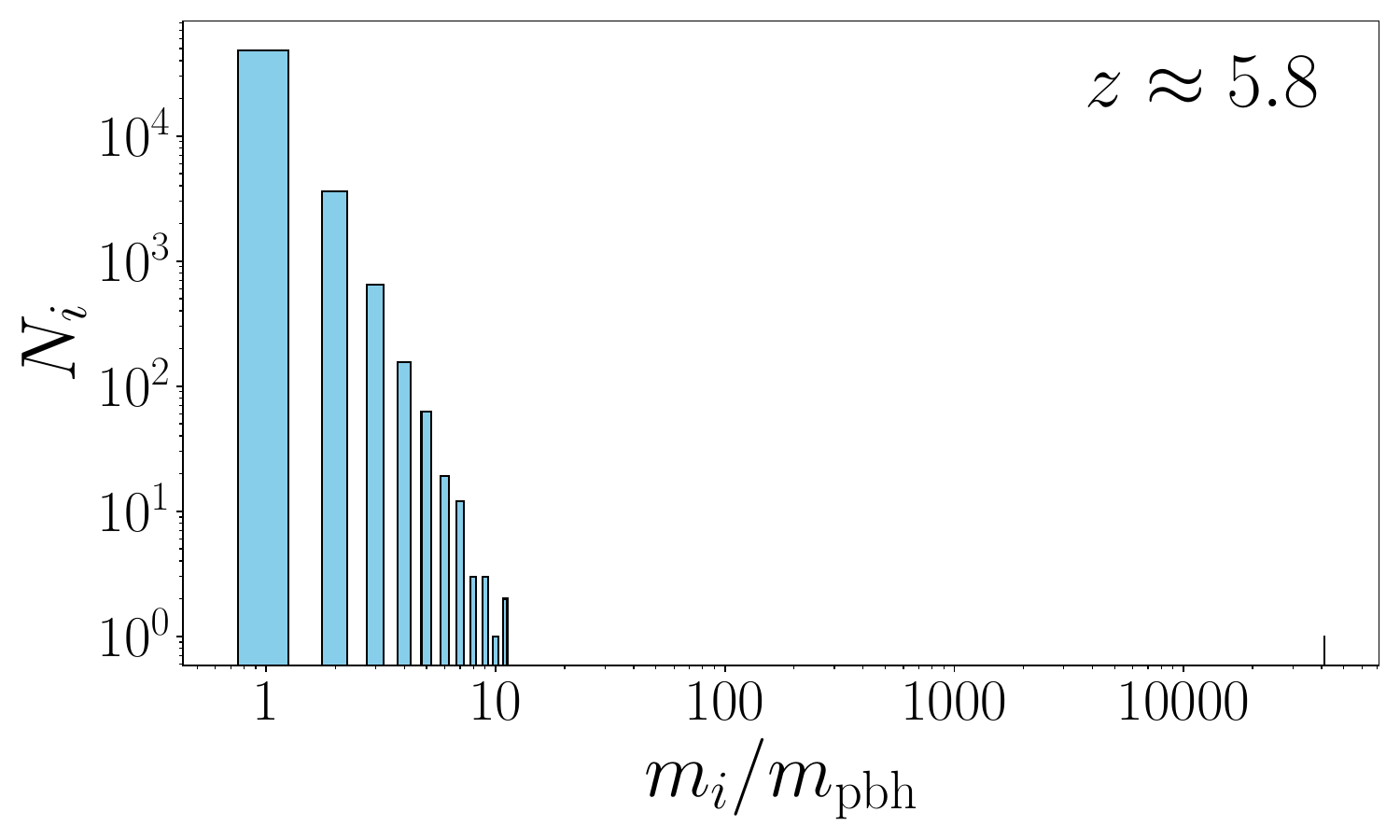}
    \label{mass_dis_4} 
    \end{minipage}
     \caption{PBH mass population evolution in a cluster of $N_{\rm cl}=10^5$ from redshift $z\simeq11.9\textup{--}5.8$ with $n_{\rm cl}=2.0\times10^8{\rm\,pc^{-3}}$ and $v_{\rm vir}=443{\rm\,km\,s^{-1}}$. The evolution of the mass distribution in a PBH cluster can be divided into three stages. During the first stage, the cluster comprises predominantly small-mass PBHs, and the merger is slow. In the second stage, intermediate-mass black holes emerge, and the merger rate increases. In the final stage, a SMBH forms, and mass increases dramatically within a short period corresponding to the runaway-merger phase. During this phase, intermediate-mass black holes merge into the central SMBH, leaving only small-mass PBHs, and an EMRI system naturally arises.}
    \label{fig:mass_distribution}
    \end{figure}

Fig.\,\ref{fig:mass_distribution} shows the evolution of PBH mass distribution for a PBH cluster with $N_{\rm cl} = 10^5$, the cluster number density $n_{\rm cl}=3N_{\rm cl}m_{\rm pbh}/4\pi r_{\rm vir}^3=2.0\times10^8{\rm\,pc^{-3}}$ and virial velocity $v_{\rm vir}=\sqrt{GN_{\rm cl}m_{\rm pbh}/r_{\rm vir}}=443{\rm\,km\,s^{-1}}$, without mass segregation, where $r_{\rm vir}$ is the cluster's virial radius. The cluster evolution can be divided into three roughly distinct stages. In the first stage, only black holes with small masses, $\sim\mathcal{O}(1)\,m_{\rm pbh}$, are present, and mergers occur slowly and fairly evenly distributed; in the second stage, black holes with intermediate masses, $\sim\mathcal{O}(10\text{--}100)\,m_{\rm pbh}$, emerge and the total merger rate increases due to the present of larger masses; in the final stage, a SMBH with mass $\gtrsim\mathcal{O}(10^3)\,m_{\rm pbh}$ forms, while the number of intermediate-mass black holes decreases as they merge into the supermassive object. At this point, the dominant process is the accretion of small PBHs—primarily first-generation objects—onto the SMBH; mergers accelerate greatly, leading to runaway growth. In a short period of time, the most massive PBH increases by several orders of magnitude, and the cluster ultimately becomes an extreme mass-ratio inspiral (EMRI) system.\,\footnote{As noted above, we do not take the result of the equation once roughly half of the cluster mass has undergone mergers. In reality, following the runaway merger phase the black holes in the cluster will not fully coalesce: some objects will be ejected from the cluster, while others remain bound and orbit the central supermassive black hole, producing extreme mass-ratio inspiral (EMRI) systems.}

\begin{figure}[htbp]
    \centering
    \begin{minipage}{0.49\linewidth}
    \centering
    \includegraphics[width=0.9\linewidth]{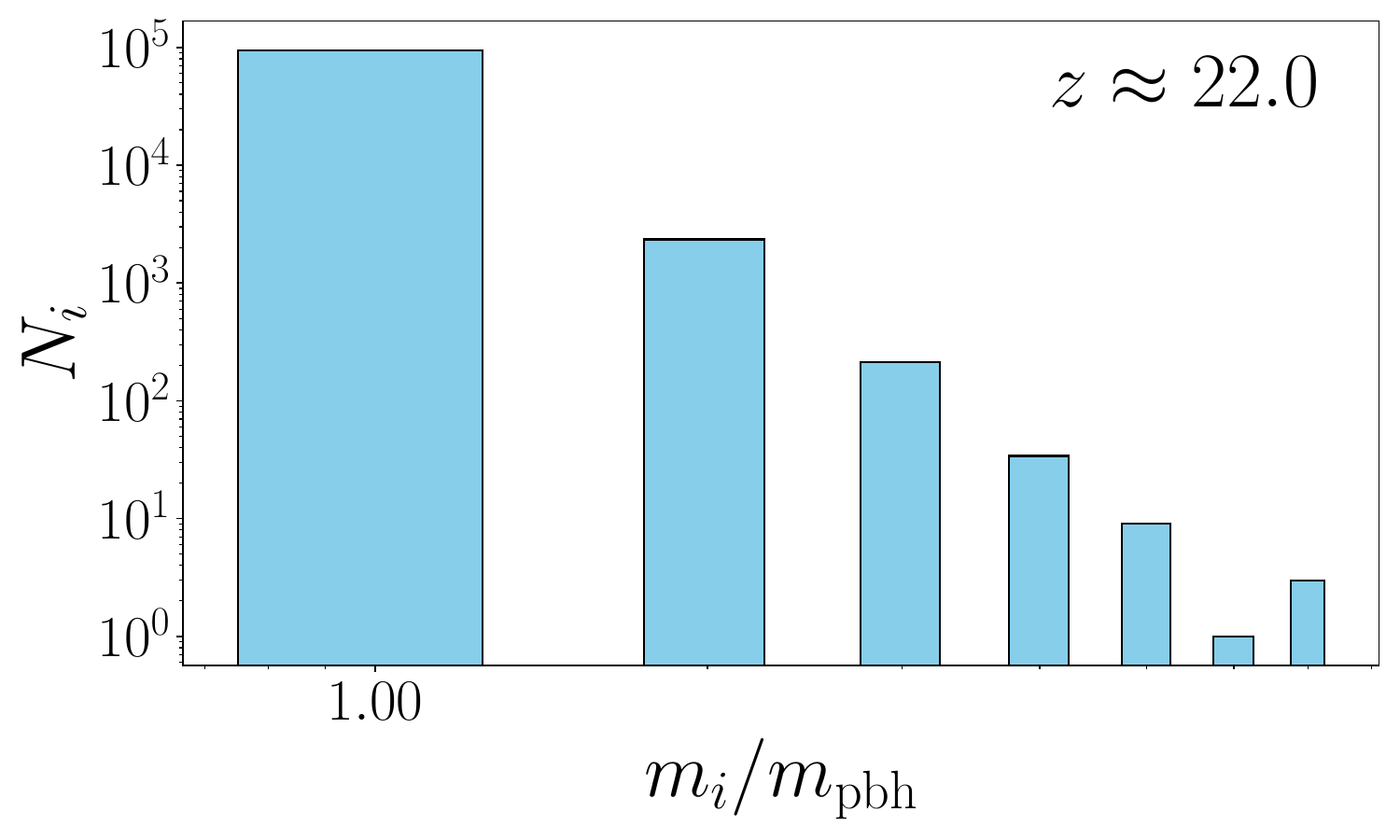}
    \label{mass_dis_1mgG} 
    \end{minipage}
    \begin{minipage}{0.49\linewidth}
    \centering
    \includegraphics[width=0.9\linewidth]{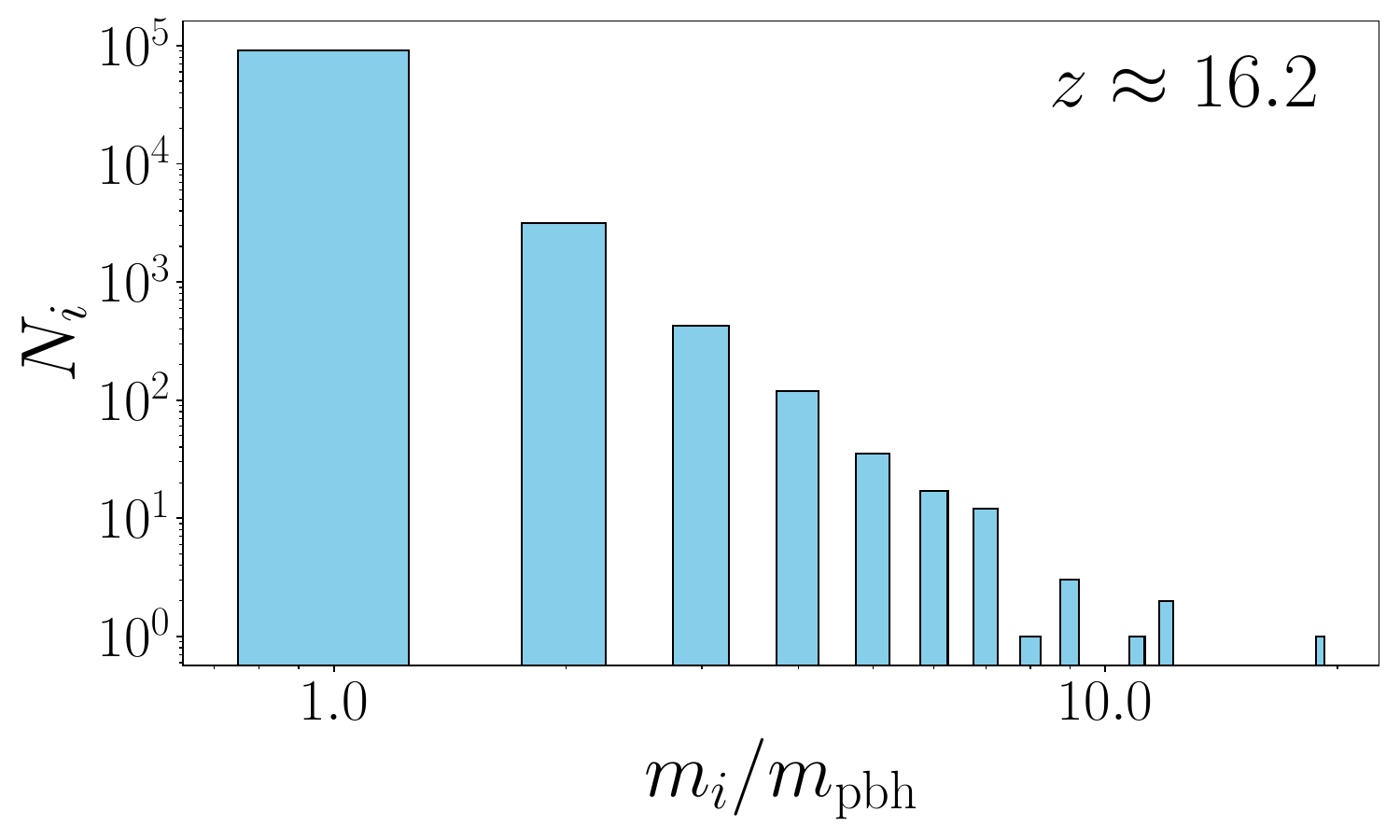}
    \label{mass_dis_2mgG} 
    \end{minipage}
    \begin{minipage}{0.49\linewidth}
    \centering
    \includegraphics[width=0.9\linewidth]{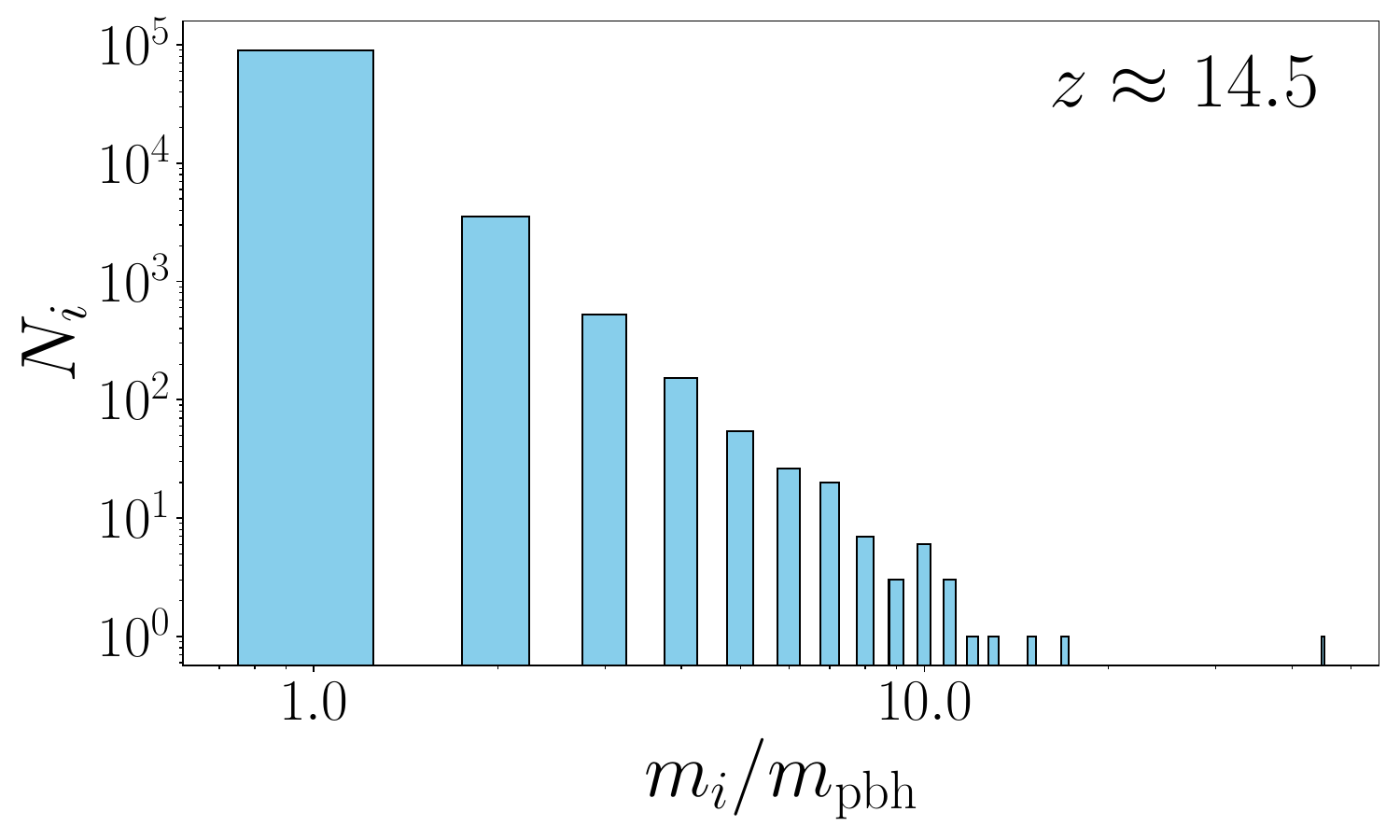}
    \label{mass_dis_3mgG} 
    \end{minipage}
    \begin{minipage}{0.49\linewidth}
    \centering
    \includegraphics[width=0.9\linewidth]{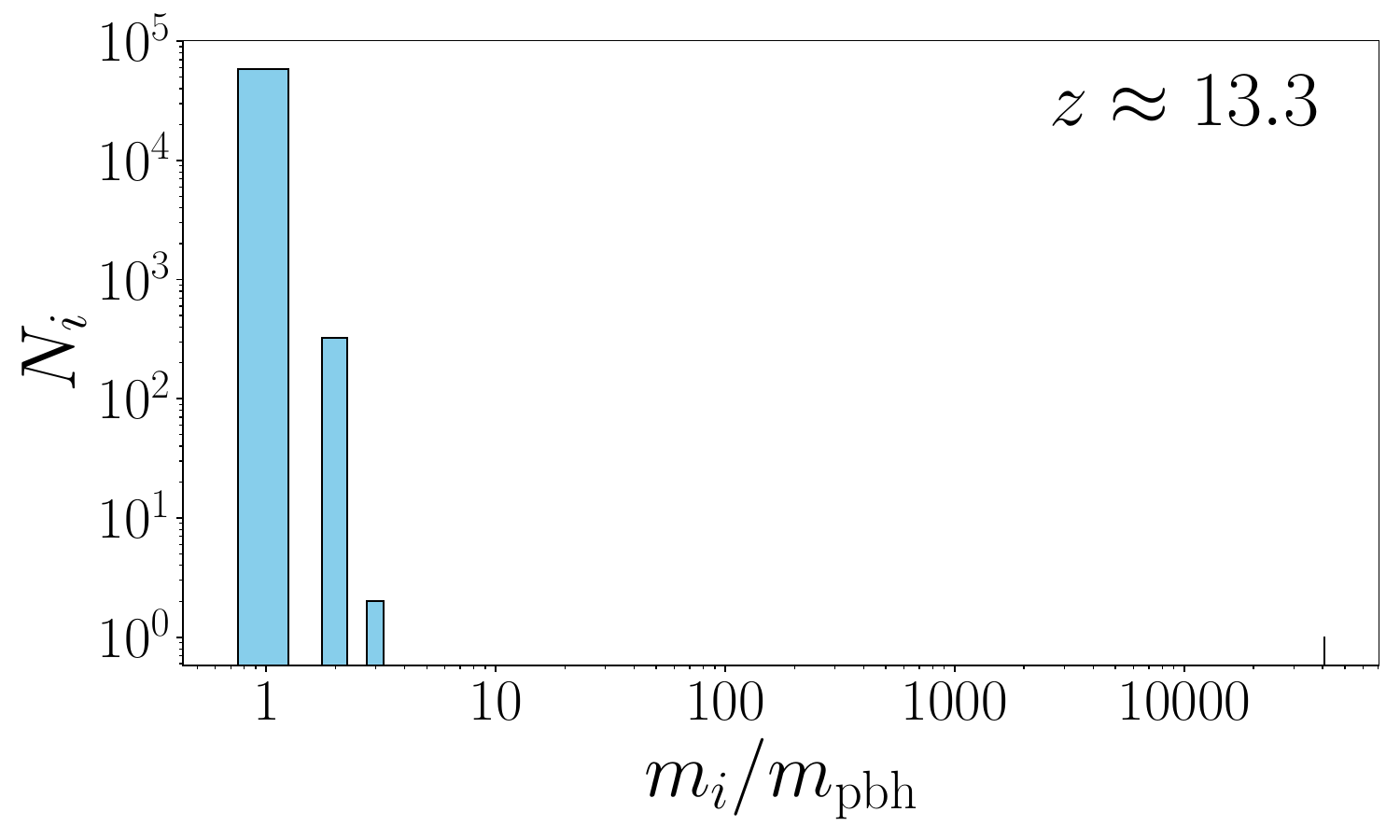}
    \label{mass_dis_4mgG} 
    \end{minipage}
     \caption{PBH mass population evolution in a cluster of $N_{\rm cl}=10^5$ from redshift $z\simeq22.0\textup{--}13.3$ with $n_{\rm cl}=2.0\times10^8{\rm\,pc^{-3}}$ and $v_{\rm vir}=443{\rm\,km\,s^{-1}}$ with mass segregation of Gaussian model.}
    \label{fig:mass_distribution_Gaussian}
    \end{figure}

\begin{figure}[htbp]
    \centering
    \begin{minipage}{0.49\linewidth}
    \centering
    \includegraphics[width=0.9\linewidth]{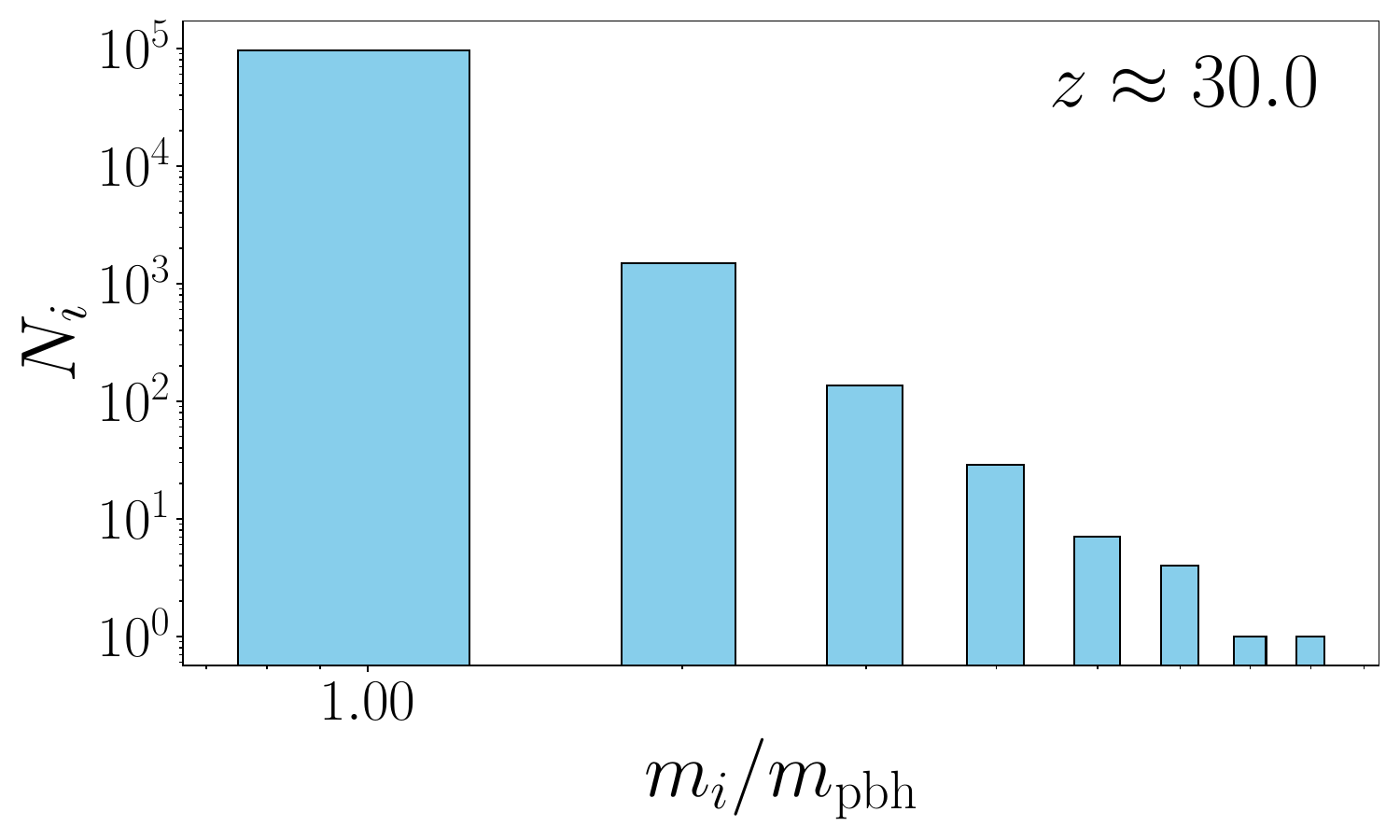}
    \label{mass_dis_1mgP} 
    \end{minipage}
    \begin{minipage}{0.49\linewidth}
    \centering
    \includegraphics[width=0.9\linewidth]{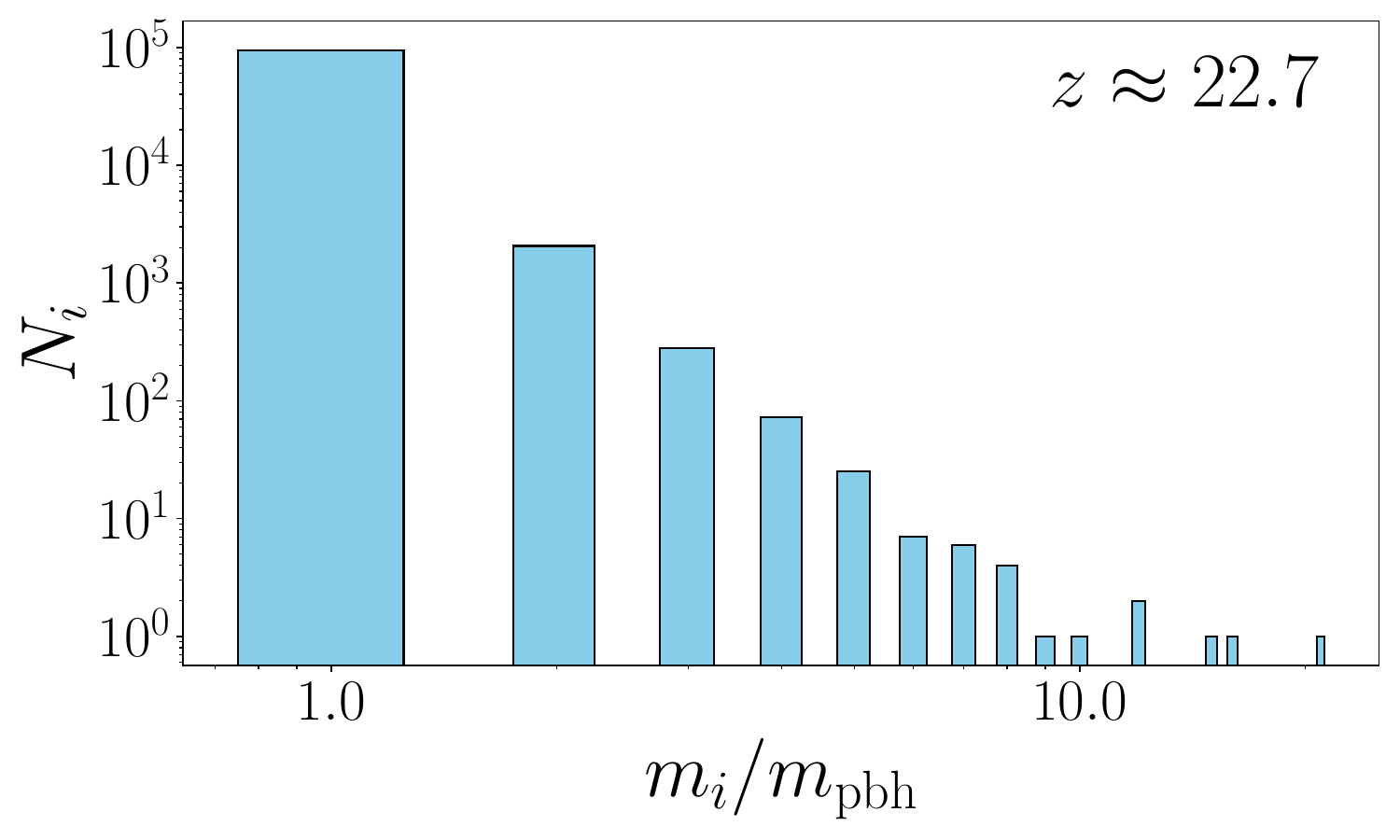}
    \label{mass_dis_2mgP} 
    \end{minipage}
    \begin{minipage}{0.49\linewidth}
    \centering
    \includegraphics[width=0.9\linewidth]{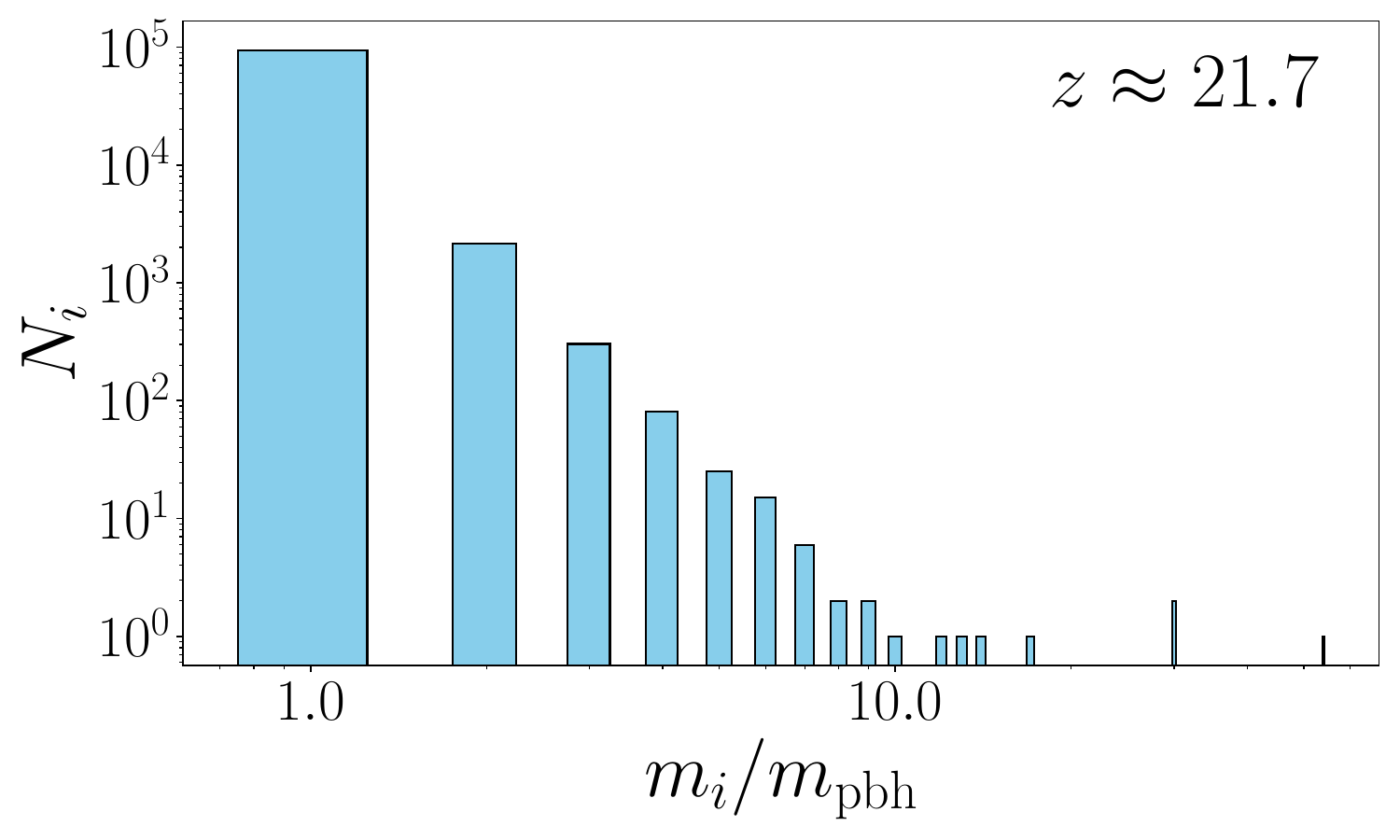}
    \label{mass_dis_3mgP} 
    \end{minipage}
    \begin{minipage}{0.49\linewidth}
    \centering
    \includegraphics[width=0.9\linewidth]{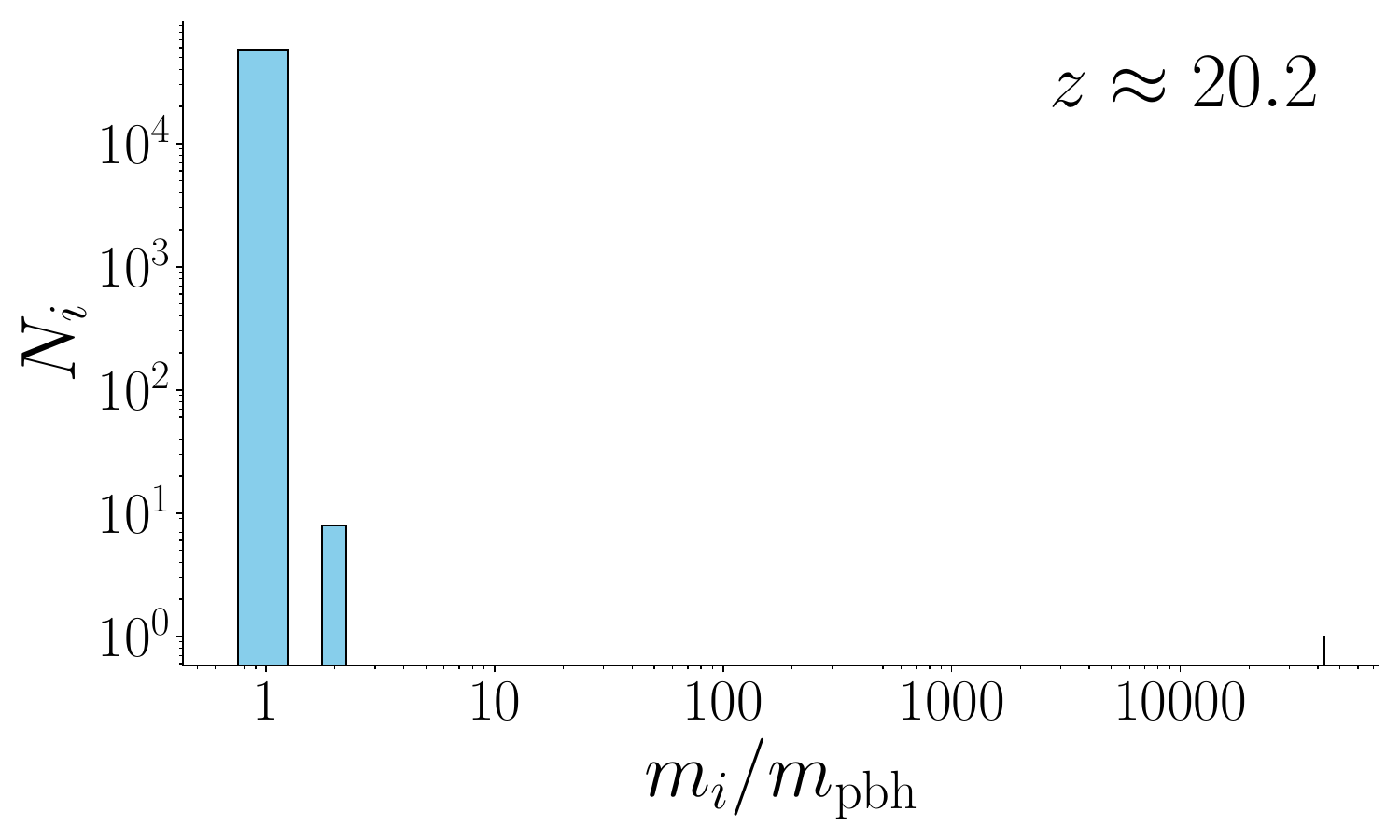}
    \label{mass_dis_4mgP} 
    \end{minipage}
     \caption{PBH mass population evolution in a cluster of $N_{\rm cl}=10^5$ from redshift $z\simeq30.0\textup{--}20.2$ with $n_{\rm cl}=2.0\times10^8{\rm\,pc^{-3}}$ and $v_{\rm vir}=443{\rm\,km\,s^{-1}}$ with mass segregation of Plummer model.}
    \label{fig:mass_distribution_Plummer}
\end{figure}

Figs.\,\ref{fig:mass_distribution_Gaussian} and \ref{fig:mass_distribution_Plummer} exhibit a simliar patern as in Fig.\,\ref{fig:mass_distribution} given the same initial cluster property with mass segregation of Gaussian and Plummer profiles, respectively, taken into account, while the runaway timescales are shortened to reach the emergence of SMBHs. In Ref.\,\cite{Zhang:2025tgm}, we demonstrate the formation of such small-scale PBH clusters requires some non-Gaussianity in the primordial power spectrum, in which we present the case without mass segregation for simplicity. With mass segregation, the emergence of SMBHs can happen even earlier at redshift $z\approx13.3$ (Gaussian) or $20.2$ (Plummer) compared to the case $z\approx5.8$ without mass segregation. In reality, mass segregation can alleviate the requirement of non-Gaussianity to form SMBHs at high redshifts. 

\subsection{Power-law scaling of PBH's number as a function of mass}
Figs.\,\ref{fig:mass_distribution}, \ref{fig:mass_distribution_Gaussian} and \ref{fig:mass_distribution_Plummer} all share a similar feature at different evolution stages. To quantify the evolution, we parametrize the cluster’s mass distribution at a fixed time by a power-law scaling:
\begin{align}
N_{i}(\tilde{t})\;\propto\;\left(\frac{m_i}{m_{\rm pbh}}\right)^{\gamma(\tilde{t})}\,,
\end{align}
where the exponent $\gamma(\tilde{t})$ is a function of evolution time $\tilde{t}$.

\begin{figure}[H]
    \centering
    \includegraphics[width=0.8\textwidth]{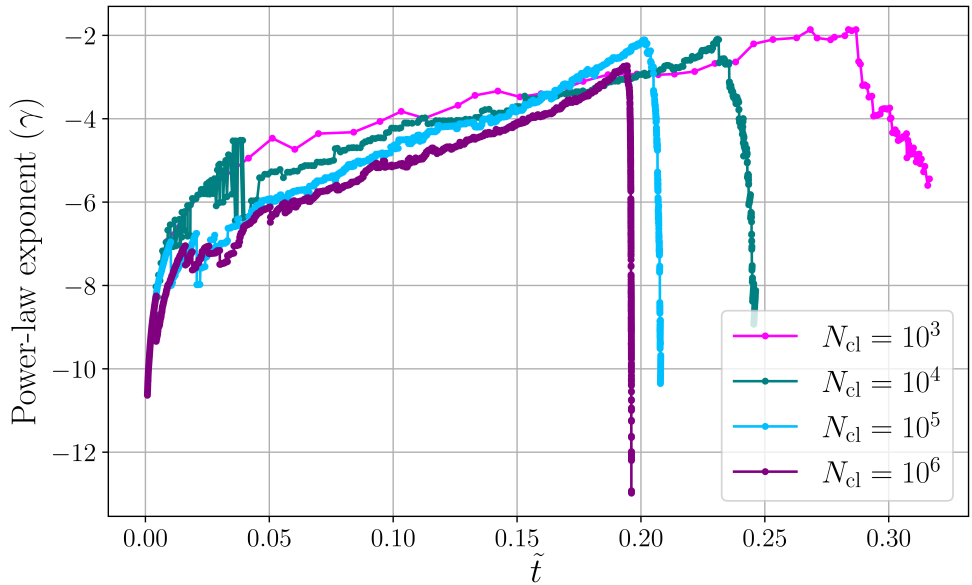}
    \caption{Time evolution of $\gamma(\tilde{t})$ for different cluster sizes $N_{\rm cl}$ for different models without mass segregation.
}\label{Fig:gammat_single}
 \end{figure}

Fig.\,\ref{Fig:gammat_single} shows the evolution of $\gamma(\tilde{t})$ for different cluster sizes, $N_{\rm cl} = 10^3, 10^4, 10^5$, and $10^6$, in the model without mass segregation. The exponent $\gamma$ rises steeply at early times, then grows approximately linearly with the dimensionless time $\tilde{t}$ over an extended interval, and finally falls rapidly near the end of the merger sequence. The initial rapid increase is caused by the early appearance of larger-mass black holes formed from first-generation mergers. The subsequent quasi-linear regime corresponds to a stage in which black holes of a broad range of masses coexist and undergo successive mergers. The abrupt decline marks the onset of the runaway-merger phase, characterized by a rapid sequence of coalescences.  
As \(N_{\rm cl}\) increases, the curves become smoother and the fluctuations decrease, as expected for a stochastic method in which statistical noise diminishes with sample size. Furthermore, in the successive-merger regime, the slope \({\rm d}\gamma/{\rm d}\tilde{t}\) is approximately independent of \(N_{\rm cl}\), indicating that the evolution of \(\gamma\) is governed primarily by the merger kernel and the structure of the Smoluchowski equation.

\begin{figure}[htbp]
    \centering
    \begin{minipage}{0.49\linewidth}
        \centering
        \includegraphics[width=0.9\linewidth]{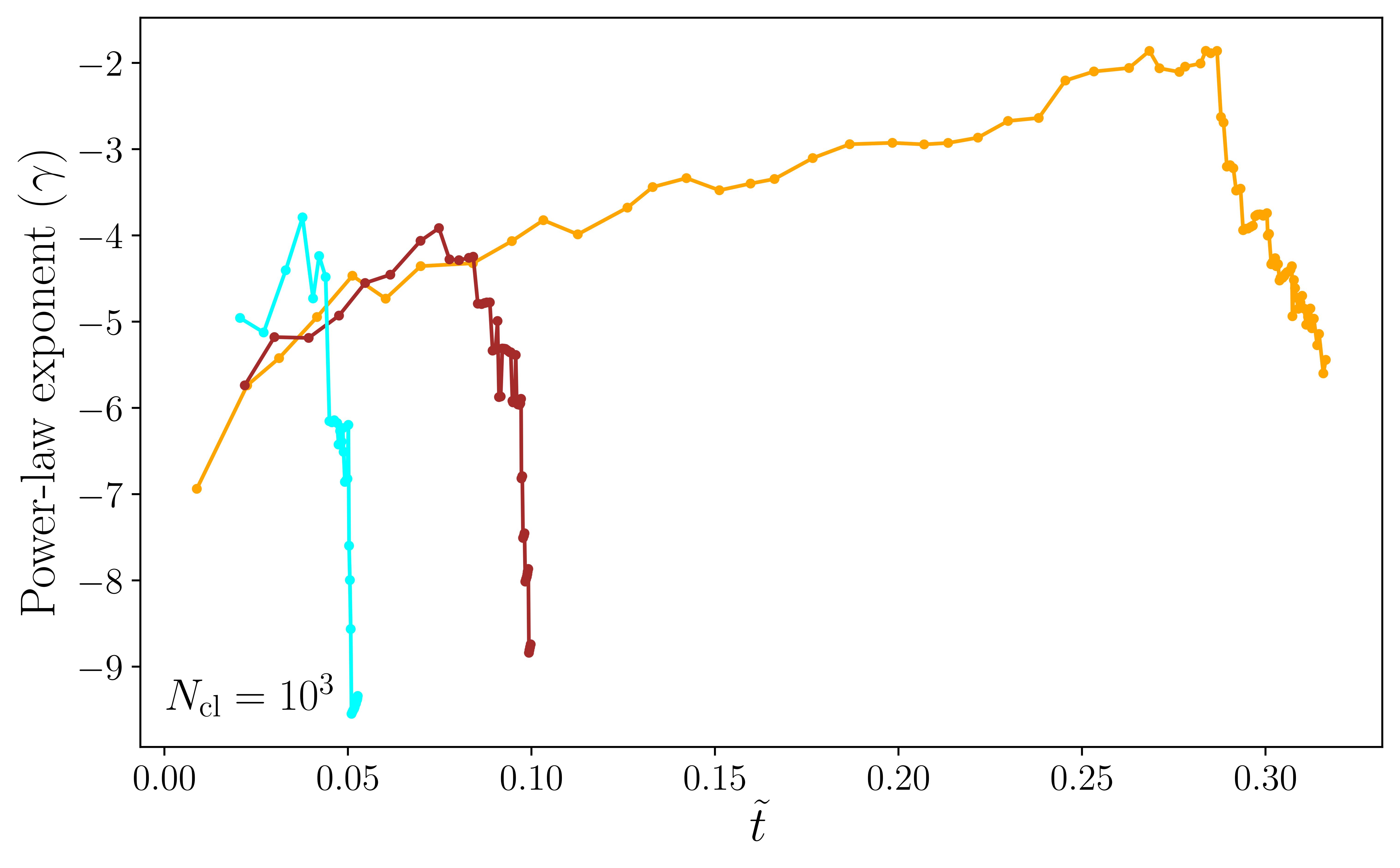}
        \label{fig:gammatN103}
    \end{minipage}
    \begin{minipage}{0.49\linewidth}
        \centering
        \includegraphics[width=0.9\linewidth]{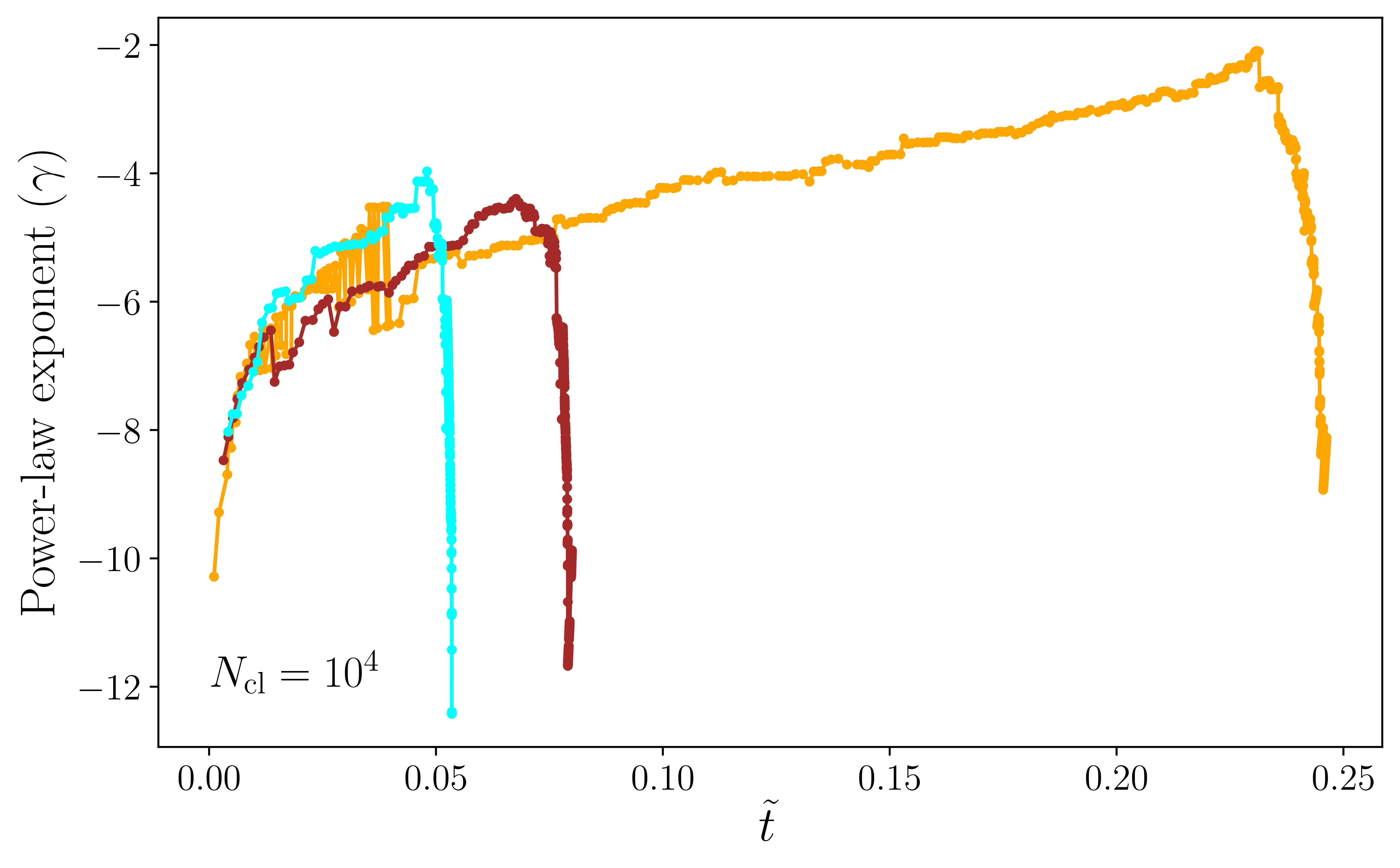}
        \label{fig:gammatN104}
    \end{minipage}

    \begin{minipage}{0.49\linewidth}
        \centering
        \includegraphics[width=0.9\linewidth]{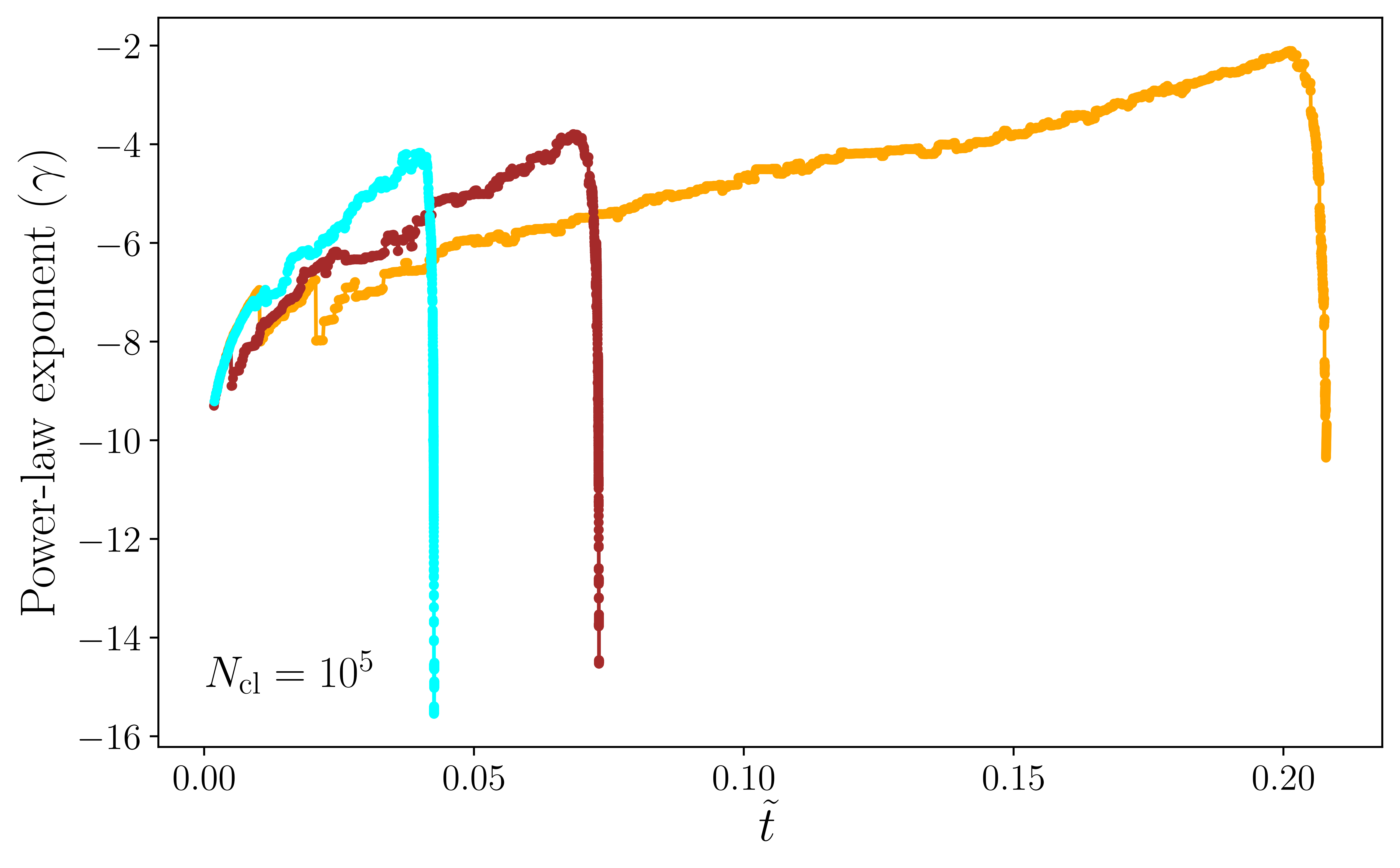}
        \label{fig:gammatN105}
    \end{minipage}
    \begin{minipage}{0.49\linewidth}
        \centering
        \includegraphics[width=0.9\linewidth]{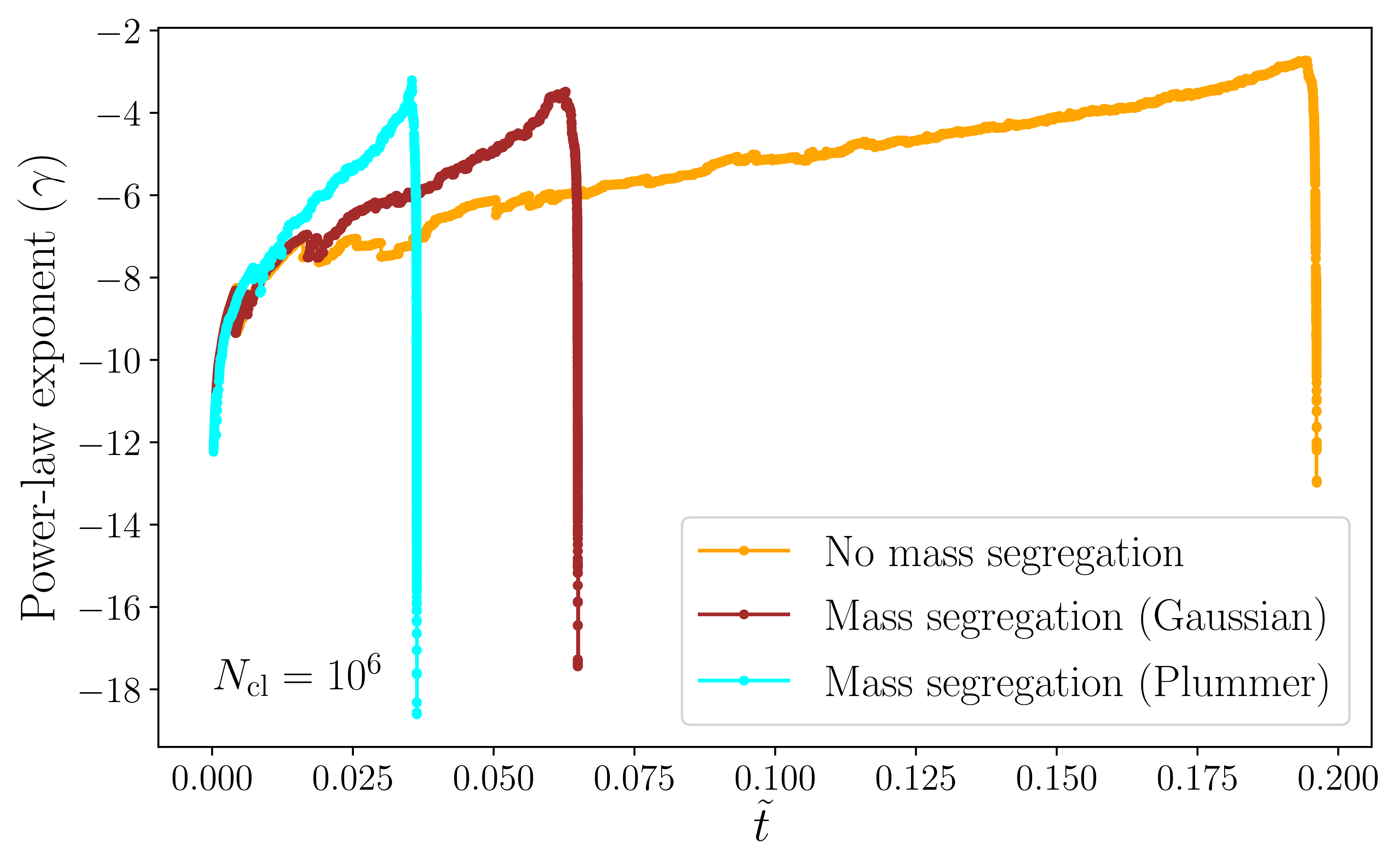}
        \label{fig:gammatN106}
    \end{minipage}

    \caption{Time evolution of $\gamma(\tilde{t})$ for different cluster sizes $N_{\rm cl}$ for different models with and without mass segregation.}
    \label{fig:gammat}
\end{figure}

Fig.\,\ref{fig:gammat} compares the evolution of $\gamma(\tilde{t})$ for models with and without (Gaussian and Plummer) mass segregation with $N_{\rm cl} = 10^3, 10^4, 10^5$, and $10^6$.
During the initial stage, the exponent $\gamma$ shows little variation between the models. However, in the successive merger regime the growth rate of $\gamma(\tilde{t})$ — the slope ${\rm d}\gamma/{\rm d}\tilde{t}$ with respect to the dimensionless runaway timescale $\tilde{t}$ — is larger (steeper) for models with mass segregation. With mass segregation, $\gamma$ attains a larger peak magnitude and subsequently declines more rapidly, $|\gamma|\sim 4$ for the Plummer and Gaussian models compared to $|\gamma|\sim2$ without mass segregation, indicating the accelerated merger process and thus the advance of the runaway merger. Moreover, the Plummer profile exhibits a faster growth of $\gamma$ and earlier runaway compared to the Gaussian model. This is consistent with the runaway-timescale calculations in the previous section, since mass segregation increases the effective merger kernel via the sinking of massive black holes toward the cluster center.

The two mass-segregation models are based on several approximations in Section\;\ref{Sec:Smoluchowski}, but generally a more accurate merger kernel can be determined by calibration. The parameters and functional forms of the kernel can be treated as free for Monte Carlo simulations in Section\;\ref{Sec:Monte_Carlo} and fit to $N$-body simulations. For example, the merger cross section—after averaging over velocity and spatial dependence—depends only on mass, so we can parameterize the mass-segregation factor as $\mathcal{F}^{\rm ms}_{ij}\propto m_i^p\,m_j^q$, or allow for a more general function $\mathcal{F}^{\rm ms}_{ij}=\mathcal{F}^{\rm ms}_{ij}(m_i,m_j)$. By fitting $p,q$ (or the full form of $\mathcal{F}^{\rm ms}_{ij}$) to $N$-body results\;\cite{Siles:2024yym} and validating the fit, we can then use those calibrated kernels in Monte Carlo simulations, which are computationally more efficient. 

\subsection{Evolution of the maximum PBH mass and the merger rate}
The characteristics of the runaway merger can be further illustrated by examining the evolution of the maximum mass in a cluster and the merger rate as functions of redshift. We calculate the merger rate as the number of mergers per unit time from our Monte Carlo simulations.

Generally, the maximum mass increases gradually over an extended period, corresponding to a stage in which only low-generation PBHs are present and no single supermassive PBH has yet formed. At later times hierarchical mergers lead to the formation of a single SMBH; the maximum mass then undergoes a short period of rapid growth, indicative of runaway behavior. The merger-rate evolution broadly follows that of the maximum mass: It increases gradually across most redshifts prior to the onset of the runaway merger. Immediately after the runaway begins, the merger rate exhibits an extremely rapid rise, and the redshift at which this peak occurs coincides with the runaway redshift of the maximum mass. Thereafter, the merger rate falls sharply owing to the depletion of surrounding low-mass PBHs. 

\begin{figure}[htbp]
\centering
\begin{minipage}{0.49\linewidth}
\centering
\includegraphics[width=0.9\textwidth]{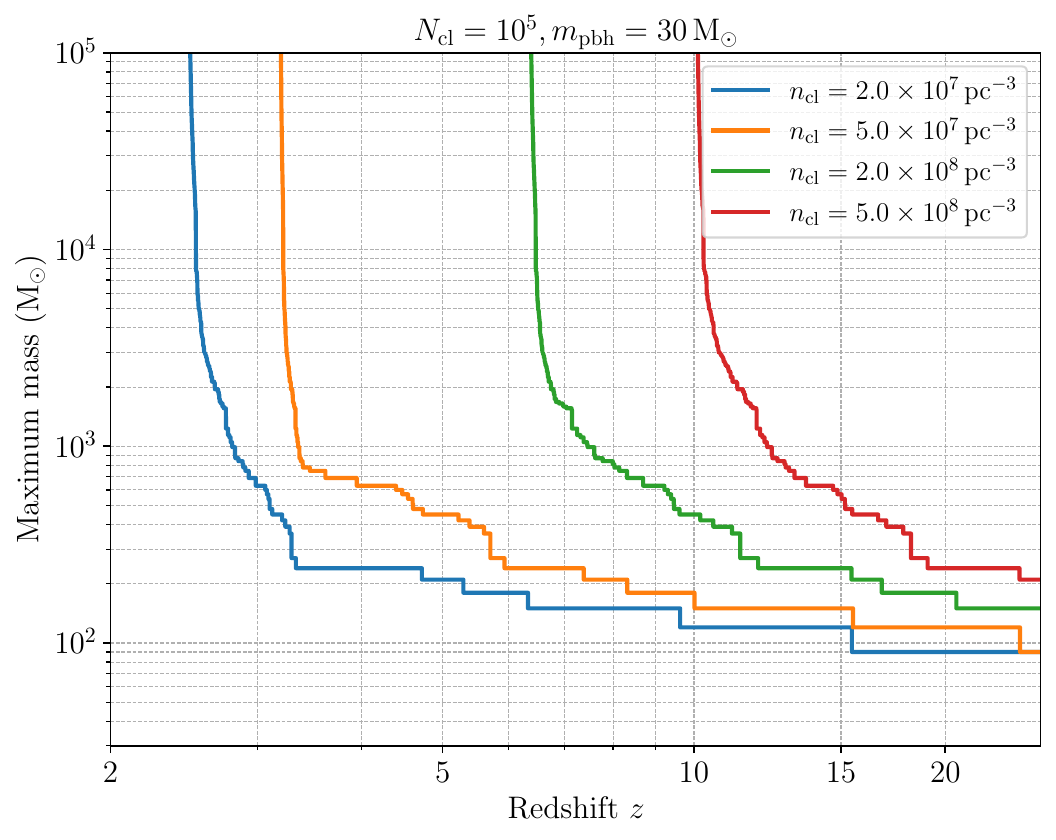}
\label{mm1} 
\end{minipage}
\begin{minipage}{0.49\linewidth}
\centering
\includegraphics[width=0.9\linewidth]{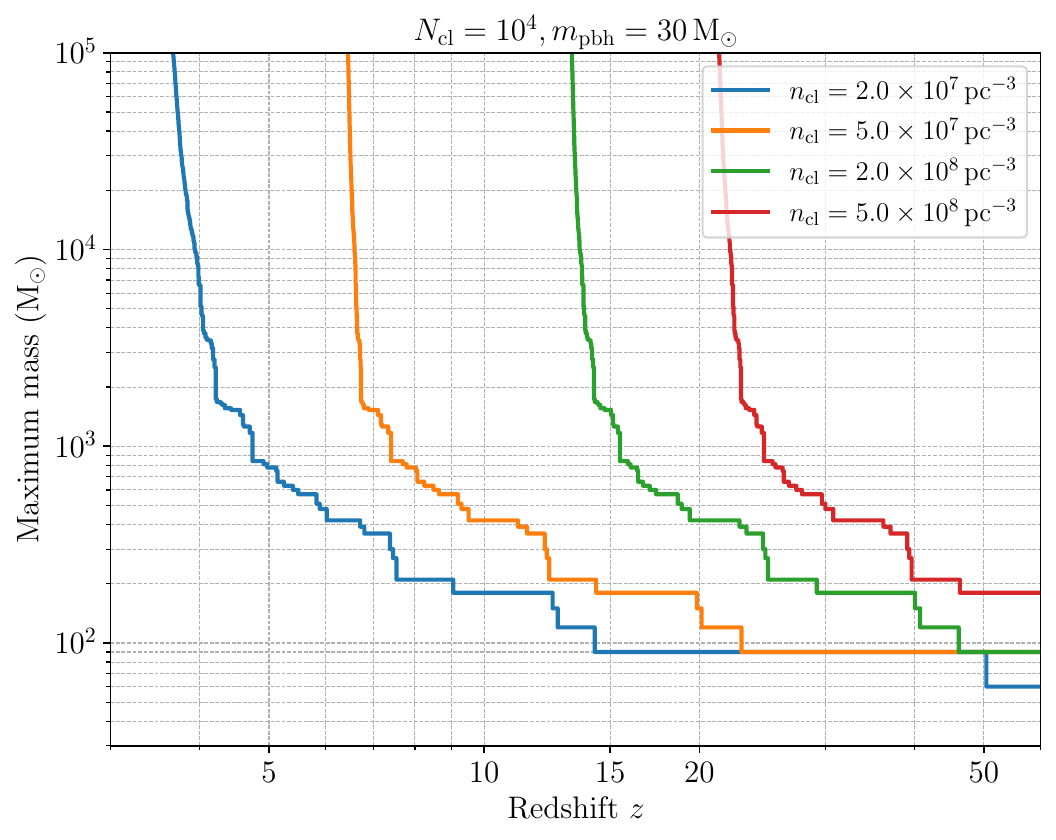}
\label{mm2} 
\end{minipage}
\begin{minipage}{0.49\linewidth}
\centering
\includegraphics[width=0.9\linewidth]{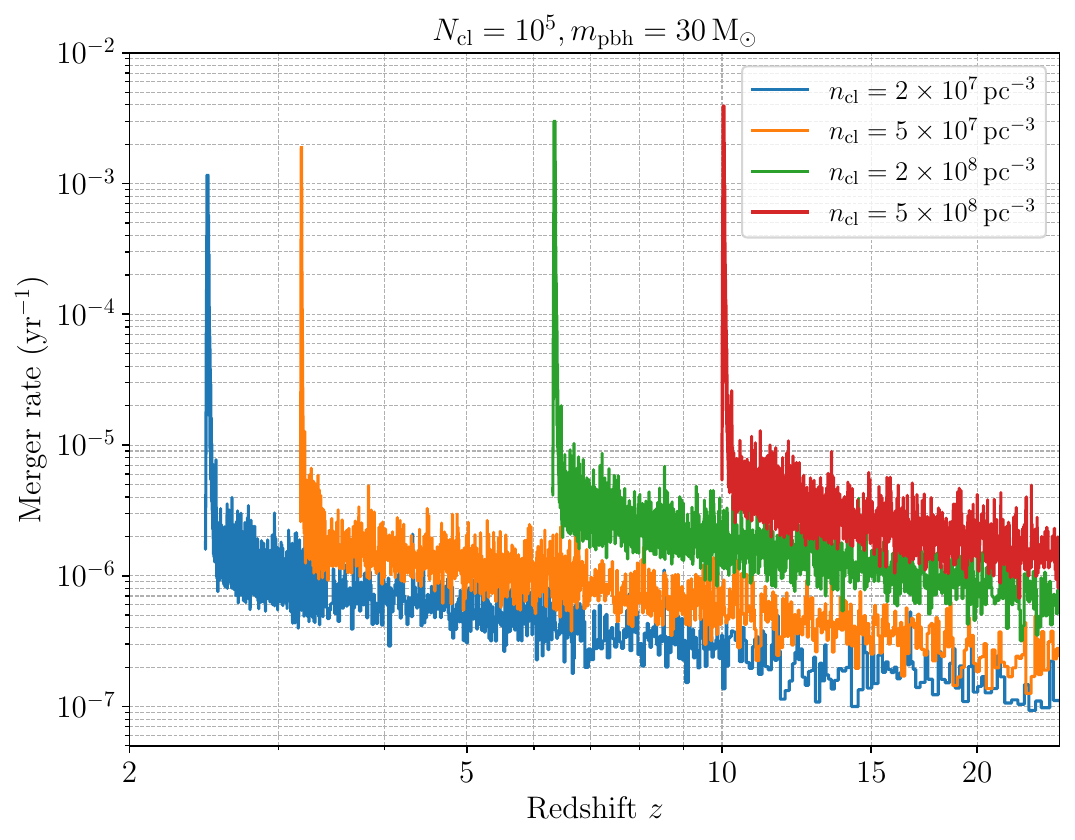}
\label{mr1} 
\end{minipage}
\begin{minipage}{0.49\linewidth}
\centering
\includegraphics[width=0.9\linewidth]{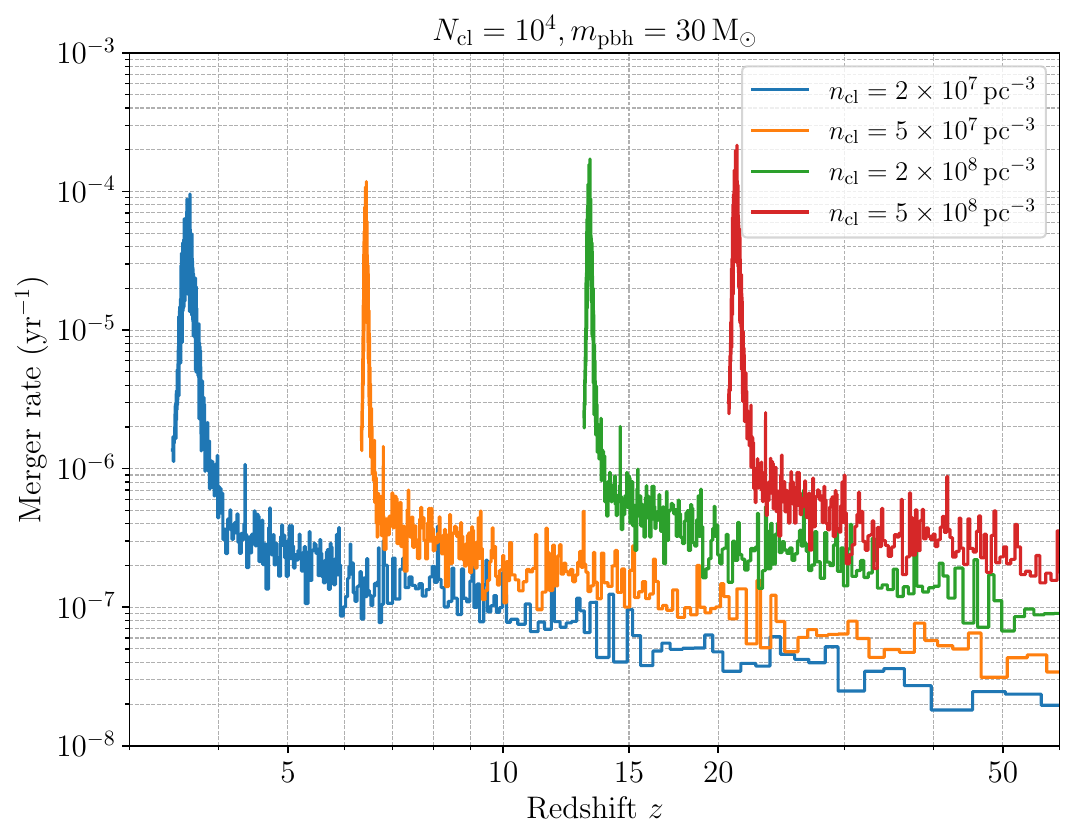}
\label{mr2} 
\end{minipage}
\caption{The evolution of the maximum PBH masses (upper panels) and the total merger rates (lower panels) in the cluster across different redshifts. The initial mass is chosen as $m_{\rm pbh}=30{\rm\,M}_{\odot}$, the total numbers are $N_{\rm cl}=10^5$ (left panels) and $N_{\rm cl}=10^4$ (right panels), respectively.}\label{Fig:maxmass}
\end{figure}

Fig.\,\ref{Fig:maxmass} demonstrates the redshift evolution of the most massive PBHs (upper panels) and the total merger rates (lower panels) in the cluster of $N_{\rm cl}=10^5$ (left panels) and $N_{\rm cl}=10^4$ (right panels) with initial PBH mass $m_{\rm pbh}=30{\rm\,M}_\odot$, respectively. For each \(N_{\rm cl}\) we vary the PBH number density $n_{\rm cl}=(2,\,5)\times10^7{\rm\,pc^{-3}}$ and $(2,\,5)\times10^8{\rm\,pc^{-3}}$ in the cluster. 
For $N_{\rm cl}=10^5\,(10^4)$ initial PBHs of mass $m_{\rm pbh}=30{\rm\,M}_\odot$ the runaway merger can occur in the matter-domination era and an SMBH forms at redshift \(z\sim 2\text{--}10\,(3\textup{--}20)\) for \(2\times10^7\,\mathrm{pc^{-3}}\lesssim n_{\rm cl}\lesssim5\times10^8\,\mathrm{pc^{-3}}\). Although a larger number density leads to a shorter runaway timescale, a smaller \(N_{\rm cl}\) results in an earlier onset of runaway merger.

This behavior can be analyzed as follows: The dimensionless runaway timescale shown in Fig.\,\ref{Fig:runaway_timescale} indicates that a larger \(N_{\rm cl}\) leads to a smaller \(\tilde{t}_{\rm ra}\), while the physical runaway timescale is given by \(t_{\rm ra}=\tilde{t}_{\rm ra}/(n_{\rm cl}\mathcal{K}_{00})\). The dependence of \(\tilde{t}_{\rm ra}\) on \(N_{\rm cl}\) is weak, approximately \(\tilde{t}_{\rm ra}\propto N_{\rm cl}^{-0.1}\), and thus the contribution of \(N_{\rm cl}\) through the virial velocity must be taken into account. Since \(r_{\rm vir}\propto n_{\rm cl}^{-1/3}N_{\rm cl}^{1/3}\) and \(v_{\rm vir}\propto N_{\rm cl}^{1/2}r_{\rm vir}^{-1/2}\propto n_{\rm cl}^{1/6}N_{\rm cl}^{1/3}\), it follows that \(1/(n_{\rm cl}\mathcal{K}_{00})\propto n_{\rm cl}^{-1}v_{\rm vir}^{11/7}\propto n_{\rm cl}^{-31/42}N_{\rm cl}^{11/21}\). Therefore, when the number density is fixed, the physical runaway timescale scales roughly as \(t_{\rm ra}\propto N_{\rm cl}^{0.42}\). In other words, a decrease of the total number of black holes shortens the runaway timescale. Physically for a fixed number density, a smaller total number of black holes implies a smaller total mass, and hence a lower virial velocity, which increases the probability of binary formation and consequently reduces the time for runaway mergers.

\section{Discussion}
\label{Sec:discussion}
Recently, JWST has discovered a new type of compact galaxies, LRDs, hosting SMBHs of $\sim10^{5\textup{--}8}{\rm\,M}_\odot$ at redshifts $z\sim4\textup{--}8$\;\cite{Harikane:2023aa,Maiolino:2023bpi,kocevski2023hidden,kokorev_uncover_2023,killi2024deciphering,kokorev_census_2024,Wang:2025ac,Durodola:2024bom,Ananna:2024jug,Kocevski:2024aa,Furtak:2024,Maiolino:2025tih,Rusakov:2026}. On the other hand, a few unexpectedly large-mass galaxies at $z\sim10$ are also reported\;\cite{Bogdan:2023ilu,Kovacs:2024zfh}. All of these call for an explanation beyond conventional astrophysics. In particular, the ``overmassive'' feature of these SMBHs\;\cite{Jones:2025aa} suggests a heavy-seed origin\;\cite{Zhang:2025oyl}. Although heavy-seed PBH of $\gtrsim10^4{\rm\,M}_\odot$ can just provide a resolution, either some non-Gaussianity in the primordial power spectrum is required or massive black holes (not quite primordial) form only after redshift $z\simeq1100$\;\cite{Qin:2025ymc,Cyr:2025tkt} in order to circumvent the CMB $\mu$-distortion constraint\;\cite{DeLuca:2025nao}. The non-Gaussianity can also produce the clustering of PBHs while satisfying all the observational constraints\;\cite{Zhang:2025tgm,Belotsky:2018wph}. If these early galaxies are seeded by PBH clusters, then our runaway-merger model could also account for the rapid emergence of SMBHs at high redshifts. 

The high spin of emergent SMBHs is another key feature of the runaway merger of PBH clustering. In Ref.\,\cite{Zhang:2025tgm}, we demonstrate that a PBH cluster can acquire spin via tidal interactions among clusters. The resulting spin can be much larger than the limit of Kerr black holes if all PBHs in the cluster were merged into an SMBH. We expect that the emergent SMBH will inherit most of the spin of the cluster\,\footnote{Even if the original cluster has zero net spin, it has been shown in full general relativity that an initial black-hole cluster of 25 equal-mass, nonspinning black holes produces a most-massive remnant with spin $\sim0.3$ times the Kerr limit~\cite{Bamber:2025gxj}.}, and the high-angular-momentum small-mass PBHs remain unmerged, surrounding the central SMBH. The EMRI system resulting from PBH cluster could potentially explain some distinct features of LRDs. For example, the V-shaped spectral energy density can be explained by a central SMBH surrounded by stellar-mass black holes\;\cite{Kritos:2025aqo,Wang:2025aa,Wang:2025ab,Chen:2026ztw}. 
In addition, X-ray quasi-periodic eruptions at the center of
galactic nuclei can be interpreted as arising from the interaction between an EMRI and the accretion disk of central SMBHs\;\cite{Kara:2025gqo}.

The Smoluchowski coagulation equation and the Monte Carlo methods employed here are not restricted to PBH clusters. They can be applied to any system governed by collisional dynamics given the appropriate coagulation kernel \(\mathcal{K}_{ij}\). For example, applying this approach to dark-halo mergers is a natural extension for future work\;\cite{Benson:2004gv,Benson:2008sd}. In addition, to obtain an analytic expression for the merger kernel we have made several assumptions and approximations; the accuracy and range of validity of this approach should therefore be tested with targeted \(N\)-body simulations.

\section{Conclusions}
\label{Sec:conclusions}

In this work, we employ a Monte Carlo method to simulate the merger evolution of a PBH cluster using the Smoluchowski coagulation equation. We compute the mass distributions for models with and without mass segregation across a range of redshifts. The mass distribution evolves through three distinct stages: (i) an initial stage dominated by mergers among  the first-generation small-mass black holes; (ii) an intermediate stage in which black holes of a broad range of masses appear and mergers among the more massive components dominate; and (iii) a final stage in which an extremely massive black hole emerges and triggers a runaway-merger phase.   

The mass distribution approximately follows a power law. Fitting the power-law exponent shows that its temporal evolution tracks the merger stages: the absolute value of the exponent decreases as new mass species appear; the system then enters a quasi-linear regime whose slope is nearly independent of the cluster size \(N_{\rm cl}\) but depends on the model (models with larger effective merger kernels exhibit larger slopes); finally, the exponent falls steeply at the onset of runaway merger. We also follow the evolution of the maximum mass in the clusters and compute the total merger rate for a range of PBH number densities and virial velocities across redshift; both quantities exhibit a clear runaway behavior. Importantly, for a plausible range of parameters, the runaway merger can occur within the first billion years, which offers a natural explanation for the mass origin of the high-redshift SMBHs recently detected by JWST.

\acknowledgments
This work is supported in part by the National Science Foundation of China (NSFC) under Grant Nos.\,12525506 and 12475107, the National Key R\&D Program of China under Grants Nos.\,2021YFC2203100 and 2017YFA0402204, the China Postdoctoral Science Foundation under Grant No.\,2024M761594, and the Shuimu Tsinghua Scholar Program.

\appendix
\numberwithin{equation}{section}

\section{The existence of runaway timescale}
\label{App:proof_runaway}
In this Section we follow the approach by Hendriks et al.\;\cite{hendriks1983coagulation} to show the existence of a runaway timescale in cases more general than those considered by Mouri and Taniguchi\;\cite{Mouri:2002mc}, including scenarios with mass segregation. A rigorous proof can be shown in the \(N\!\to\!\infty\) limit; this result provides an upper bound for finite-size systems. Here we show that if the merger kernel admits a lower bound given by a convex, positively homogeneous function of degree $\lambda$, i.e.,
$\tilde{\mathcal{K}}(s i, s j)=s^{\lambda}\tilde{\mathcal{K}}(i,j)$ for $s>0$, then \(\lambda>1\) implies the existence of a finite runaway timescale.

We first introduce the partial moments\;\cite{hendriks1983coagulation}:
\begin{align}
\tilde{ M}_{\alpha,Q}(t)=\sum^Q_{i=1}i^\alpha \tilde{n}_i(t)\;.
\end{align}
They satisfy the evolution equations:
\begin{align}\label{Eq:finite_moment_eq}
\dot{\tilde{M}}_{\alpha, Q}=\frac{1}{2}\sum^{Q-1}_{i=1}\sum^{Q-i}_{j=1}\tilde{n}_i\tilde{n}_j\tilde{\mathcal{K}}_{ij}\left((i+j)^\alpha-i^\alpha-j^\alpha\right)-\sum^Q_{i=1}\sum^{\infty}_{j=Q-i+1}\tilde{n}_i\tilde{n}_j\tilde{\mathcal{K}}_{ij}i^\alpha\,,
\end{align}
as derived from the dimensionless Smoluchowski equation, Eq.\,\ref{Eq:Smol_dimless} in the \(N\to\infty\) limit, by multiplying with \(i^{\alpha}\), summing over \(i\), and rearranging terms, where a dot denotes a time derivative, i.e. \(\dot{\tilde M}_{\alpha,Q}\equiv {\rm d}\tilde M_{\alpha,Q}/{\rm d}t\).
Furthermore, it can be shown that if
\begin{align}
\label{Eq:S(Q)}
\lim_{Q\to\infty}\sum_{i=1}^{Q} \tilde{n}_i\,\tilde{\mathcal{K}}_{ij}\,i^{\alpha} < \infty
\quad \text{for all fixed } j\in\mathbb{N},
\end{align}
then (we will prove this at the end of this Section)
\begin{align}
\label{Eq:T(Q)}
\lim_{Q\to\infty}\sum_{i=1}^{Q}\sum_{j=Q-i+1}^{\infty}
\tilde{n}_i\,\tilde{n}_j\,\tilde{\mathcal{K}}_{ij}\,i^{\alpha} = 0.
\end{align}
Therefore, the moment equation becomes the following:
\begin{align}
\dot{\tilde{M}}_{\alpha}=\frac{1}{2}\sum^\infty_{i=1}\sum^\infty_{j=1}\tilde{n}_i\tilde{n}_j\tilde{\mathcal{K}}_{ij}\left((i+j)^\alpha-i^\alpha-j^\alpha\right)
\end{align}
as $Q\to\infty$. Clearly, the first ($\alpha=1$) moment equation, $\dot{\tilde{M}}_1(t)=0$, meaning the conservation of total mass, $\tilde{M}_1\equiv1$. While the zeroth ($\alpha=0$) moment equation,
\begin{align}
\label{Eq:zero_moment}
\dot{\tilde{M}}_0=-\frac{1}{2}\sum_{ij}\tilde{n}_i\tilde{n}_j\tilde{\mathcal{K}}_{ij}<0\,,
\end{align}
implying the total number density $\tilde{M}_0(t)$ is monotonically decreasing during evolution, and the second ($\alpha=2$) moment equation,
\begin{align}
\label{Eq:second_moment}
\dot{\tilde{M}}_2=\sum_{ij}\tilde{n}_i\tilde{n}_j\tilde{\mathcal{K}}_{ij}ij>0\,,
\end{align}
describing the growth of average mass $\tilde{M}_2/\tilde{M}_1=\tilde{M}_2(t)$. We consider a class of merger kernels that are positive homogeneous functions of degree $\lambda$, that is,
\begin{align}
\tilde{\mathcal{K}}(si,sj)=s^\lambda\tilde{\mathcal{K}}(i,j)\;.
\end{align}
A useful tool to obtain the upper bound of the runaway timescale is the Jensen's inequality\;\cite{Jensen:1906}: Given a set of positive numbers $c_k$ and a convex function $\tilde{\mathcal{K}}(x)$, the expectation values of any set $b_k$ is  
\begin{align}
\mathbf{E}(b)=\frac{\sum b_kc_k}{\sum c_k}\,,
\end{align}
and it satisfies the inequality,
\begin{align}
\mathbf{E}\left(\tilde{\mathcal{K}}(b)\right)\geq \tilde{\mathcal{K}}\left(\mathbf{E}(b)\right)\;.
\end{align}

We then apply this inequality to Eq.\,\ref{Eq:second_moment}. The expectation value of $\tilde{\mathcal{K}}(i,j)$ satisfies
\begin{align}
\mathbf{E}(\tilde{\mathcal{K}}(i,j))_j=\frac{\sum_{i}i\tilde{n}_i\tilde{\mathcal{K}}(i,j) }{\sum_i i\tilde{n}_i}
\geq \tilde{\mathcal{K}}\left(\frac{\sum_i i^2\tilde{n}_i}{\sum_ii\tilde{n}_i},j\right)
=\tilde{\mathcal{K}}\left(\frac{\tilde{M}_2}{\tilde{M}_1},j\right)\,,
\end{align}
where the subscript $j$ on $\mathbf{E}(\tilde{\mathcal{K}}(i,j))$ denotes calculation under fixed $j$.
Thus
\begin{align}
\dot{\tilde{M}}_2&=\sum_{j}j\tilde{n}_j\sum_i\tilde{n}_i\tilde{\mathcal{K}}_{ij}i=\left(\sum_i i\tilde{n}_i\right)\sum_{j}j\tilde{n}_j\mathbf{E}(\tilde{\mathcal{K}}(i,j))_j=\tilde{M}_1\sum_{j}j\tilde{n}_j\mathbf{E}(\tilde{\mathcal{K}}(i,j))_j\notag\\
&\geq\tilde{M}_1\sum_{j}j\tilde{n}_j\tilde{\mathcal{K}}\left(\frac{\sum_i i^2\tilde{n}_i}{\sum_ii\tilde{n}_i},j\right)\geq \tilde{M}_1^2 \tilde{\mathcal{K}}\left(\frac{\sum_i i^2\tilde{n}_i}{\sum_ii\tilde{n}_i},\frac{\sum_j j^2\tilde{n}_j}{\sum_jj\tilde{n}_j}\right)\;.
\end{align}
Since the kernel considered here is a homogeneous function, we obtain
\begin{align}
\dot{\tilde{M}}_2\geq \tilde{M}_1^2\tilde{\mathcal{K}}\left(\frac{\tilde{M}_2}{\tilde{M}_1},\frac{\tilde{M}_2}{\tilde{M}_1}\right)=\tilde{\mathcal{K}}(1,1)\tilde{M}_2^\lambda=\tilde{M}_2^\lambda\,,
\end{align}
where we have used $\tilde{\mathcal{K}}(1,1)=1$ by definition and the conservation of total mass, $\tilde{M}_1 = 1$. In our case, we focus on the degree $\lambda > 1$. After performing the integration, we obtain the corresponding lower bound for $\tilde{M}_2$.
\begin{align}
\tilde{M}_2(\tilde{t})\geq \tilde{M}_2(0)\left(\frac{1}{1-\tilde{t}/\tilde{t}_{2,c}}\right)^{\frac{1}{\lambda-1}}.
\end{align}
For $\lambda>1$, the average mass $\tilde{M}_2/\tilde{M}_1=\tilde{M}_2(t)$ diverges at $\tilde{t}_{2,c}=\left(|\lambda-1|\tilde{M}_2^{\lambda-1}(0)\right)^{-1}$, due to the emergence of infinite-mass particles thus the violation of mass conservation, $\dot{\tilde{M}}_1\neq0$\;\cite{hendriks1983coagulation}. As our proof is based on the dimensionless Smoluchowski equation, the result corresponds to the upper bound of the dimensionless runaway timescale. Consequently, the upper bound of the physical runaway timescale is given by
\begin{align}
t= \frac{\tilde{t}}{n_0\mathcal{K}_{00}}\leq  \frac{\tilde{t}_{2,c}}{n_0\mathcal{K}_{00}}=\frac{1}{n_0\mathcal{K}_{00}|\lambda-1|}\,,
\end{align}
where we have used $\tilde{M}_2(0)=1$. The equation above is consistent with the result derived in Ref.\,\cite{Mouri:2002mc}. However, our approach derives the upper bound in a more general manner, which can be applied to broader cases.

\subsection*{Proof of Eq.\,\ref{Eq:T(Q)} given Eq.\,\ref{Eq:S(Q)}}
The proof is divided into three steps:
\begin{enumerate}
\item Define, for fixed \(j\),
\begin{align}
S_j(Q)&\equiv\sum_{i=1}^{Q}\tilde n_i\,\tilde{\mathcal K}_{ij}\,i^{\alpha}\,,
\end{align}
and
\begin{align}
T(Q)&\equiv \sum_{i=1}^{Q}\sum_{j=Q-i+1}^{\infty}\tilde n_i\,\tilde n_j\,\tilde{\mathcal K}_{ij}\,i^{\alpha}
= \sum_{j=1}^{\infty}\sum_{i=Q-j+1}^{Q}\tilde n_i\,\tilde n_j\,\tilde{\mathcal K}_{ij}\,i^{\alpha}\,,
\end{align}
where the equality follows from reordering the (nonnegative) summation indices. We obtain
\begin{align}
T(Q)=\sum_{j=1}^{\infty} \tilde n_i\bigl(S_j(Q)-S_j(Q-j)\bigr)\;.
\end{align}
\item Assume that for each fixed \(j\) the sequence \(S_j(Q)\) has a finite limit,
\begin{align}
S_j(\infty):=\lim_{Q\to\infty}S_j(Q)<\infty\;.
\end{align}
Then for each fixed \(j\),
\begin{align}
\lim_{Q\to\infty}\bigl(S_j(Q)-S_j(Q-j)\bigr)=S_j(\infty)-S_j(\infty)=0\,,
\end{align}
and hence
\begin{align}
\lim_{Q\to\infty}\tilde{n}_j\bigl(S_j(Q)-S_j(Q-j)\bigr)=0
\quad\text{for every fixed }j\;.
\end{align}
\item To interchange the limit and the infinite sum in $T(Q)$, we apply the Lebesgue dominated convergence theorem. Given that
\begin{align}
\big|\tilde{n}_j\left(S_j(Q)-S_j(Q-j)\right)\big|\leq g_j\,,
\end{align}
for all large $Q$, and $\sum_jg_j<\infty$, we can interchange the sum and the limit. 
It is easy to check that $S_j$ satisfies these conditions:
\begin{align}
\big|\left(S_j(Q)-S_j(Q-j)\right)\big|\leq S_j(Q)\leq S_{j}(\infty)
\end{align}
as $S_j(Q)$ is increasing with $Q$ (with all terms positive). We can then choose $g_j=\tilde{n}_jS_j(\infty)$ so that
\begin{align}
\big|\tilde{n}_j\left(S_j(Q)-S_j(Q-j)\right)\big|\leq\tilde{n}_jS_j(\infty)\;.
\end{align}
Thus
\begin{align}
T(Q)=\lim_{Q\to\infty}\sum_{j=1}^\infty \tilde{n}_j\left(S_j(Q)-S_j(Q-j)\right)=\sum_{j=1}^{\infty}\lim_{Q\to\infty}\tilde{n}_j\left(S_j(Q)-S_j(Q-j)\right)=\sum_{j=1}^\infty 0=0\,,
\end{align}
which gives the desired result. Given that at least one of $\tilde{n}_j$ is nonzero (nonempty system), we can also verify the assumption, Eq.\,\ref{Eq:S(Q)}, from
\begin{align}
\sum_j^\infty \tilde{n}_jS_j(\infty)=\sum_j^\infty\sum_i^\infty \tilde{n}_j\tilde{n}_i\tilde{\mathcal{K}}_{ij}i^\alpha <\infty\,,
\end{align}
which should hold if the system is physical as it is the same type of double sum in the first term of Eq.\,\ref{Eq:finite_moment_eq} in the limit $Q\to \infty$.
\end{enumerate}

\section{Sampling methods}
\label{App:Sampling methods}
\subsection*{The acceptance-rejection method}\label{Append:ARS}
In many Monte Carlo applications, one needs to sample from a target probability density \(f(x)\) which may be difficult to invert directly.  The acceptance-rejection method provides a simple way to draw independent samples from \(f\).  One chooses an “envelope” density \(g(x)\), from which sampling is easy, and a constant \(c\) such that
\begin{align}
    f(x) \le c\,g(x)\quad\forall x.
\end{align}
The acceptance-rejection algorithm proceeds as follows:
\begin{enumerate}
  \item Draw \(X\sim g(x)\) and \(U\sim \mathrm{Uniform}(0,1)\), independent.
  \item Compute the acceptance probability
    \begin{align}
      P_{\rm acc} = \frac{f(X)}{c\,g(X)}.
    \end{align}
  \item If \(U \le P_{\rm acc}\), \textbf{accept} \(X\) as a draw from \(f\); otherwise, \textbf{reject} and return to step 1.
\end{enumerate}

\noindent
It can be shown that the accepted \(X\) indeed follow the desired density \(f(x)\).  The overall efficiency of the method is
\begin{align}
  \Pr(\textbf{accept}) = \int \! g(x)\,\frac{f(x)}{c\,g(x)}\,\mathrm{d}x
  = \frac{1}{c}\,,
\end{align}
so one typically chooses \(g\) and \(c\) to minimize \(c\), balancing ease of sampling against closeness to \(f\).  

\subsection*{The inversion transformation method}\label{Append:inversion transformation method}
The inverse transform sampling method is a fundamental technique for generating random variables from an arbitrary continuous distribution whose cumulative distribution function (CDF) \(F(x)\) is known.  The core observation is that if
\begin{align}
    U \sim \mathrm{Uniform}(0,1),
\end{align}
then the variable
\begin{align}
    X = F^{-1}(U)
\end{align}
has a distribution function \(F\).  Here, the \textbf{generalized inverse} \(F^{-1}: [0,1]\to\mathbb{R}\) is defined by
\begin{align}
    F^{-1}(u) \,=\, \inf\{x\in\mathbb{R}: F(x) \ge u\}.   \label{eq:gen-inv}
\end{align}

\paragraph{Derivation.}
Under the assumption that \(F\) is continuous and strictly increasing, one can show directly that
\begin{align*}
    \Pr(X \le x) 
    &= \Pr\bigl(F^{-1}(U) \le x\bigr)
    = \Pr\bigl(U \le F(x)\bigr)
    = F(x).
\end{align*}
When \(F\) has flat regions (as in mixed or discrete components), the generalized inverse definition~\eqref{eq:gen-inv} still ensures \(X\) has the correct distribution.

\paragraph{Algorithm.}  A succinct description of the sampling procedure:
\begin{enumerate}
  \item Draw \(U \sim \mathrm{Uniform}(0,1)\).
  \item Compute \(X = F^{-1}(U)\) (using an analytic expression or numerical inversion).
\end{enumerate}

\paragraph{Implementation Details.}
\begin{itemize}
  \item \textbf{Analytic inverse.}  For many standard distributions (e.g., exponential, logistic, Pareto, Weibull), closed-form inverses exist.  These lead to {exact} and {efficient} samplers requiring only one uniform draw per variate.
  \item \textbf{Numerical root-finding.}  When \(F^{-1}\) is not explicit, one can solve \(F(x) = u\) for \(x\) via methods such as bisection, Newton--Raphson, or secant methods.  Choose bracketed methods (e.g., bisection) for guaranteed convergence in monotonic contexts.
  \item \textbf{Interpolation.}  If \(F\) is available only as a lookup table or empirical CDF, one may apply linear or spline interpolation to approximate the inverse.  Ensure monotonicity by enforcing non-decreasing interpolants.
 \item \textbf{Discrete distributions.}  For purely discrete distributions with probabilities \(p_i = \Pr(X = x_i)\), form the stepwise CDF
  \(F(x_i) = \sum_{j\le i} p_j\), draw \(U\), and select the smallest \(x_i\) with \(F(x_i) \ge U\).
\end{itemize}

\begin{description}
  \item[Advantages:]  Exactness (modulo numerical error), simplicity, and generality.  Easy to implement once an inverse CDF is known.
  \item[Limitations:]  Requires computing \(F^{-1}\).  Numerical inversion can be expensive if \(F\) is costly to evaluate.  Mixed or heavy-tailed distributions may require careful handling of tail regions or piecewise strategies.
\end{description}

\paragraph{Example:}
For the exponential distribution with rate \(\lambda>0\),
\begin{align*}
    F(x) = 1 - e^{-\lambda x}, \quad x\ge0,
\end{align*}
one solves \(F(x) = u\) by
\begin{align*}
    x = -\frac{1}{\lambda} \ln(1-u),
\end{align*}
and since \(1-U\) is also uniform on \([0,1]\), one often writes
\(X = -\lambda^{-1} \ln U\).

\subsection*{Discrete inversion transformation method} In the above, we mainly discuss the case that the distribution function is continuous, we will give more details of the discrete case since in our case for simulating merger of PBHs, the probability density function $f$ is defined with respect to a discrete variable $i$ ranging over all integers between $m$ and $n$, we should replace $x$ by $i$ and integrals by sums. And $f(i_0),m\leq i_0\leq n$ represents the probability that the random integer $i$ equals to $i_0$. The CDF is defined as
\begin{align}
    F(i)=\sum_{i^\prime=m}^if(i^\prime),\quad m\leq i\leq n
\end{align}
which represents the probability that $i$ is less than or equal to $i$.

The key point is to generate a random integer $i$ according to the probability distribution, the method is straightforward, unlike the continuous case, we select $i$ as the variable that satisfies
\begin{align}
    F(i-1)<r\leq F(i) 
\end{align}
where $r$ is a random number from a uniform distribution $r\in[0,1]$. The normalization condition should be satisfied:
\begin{align}
    \sum_{i=m}^nf(i)=1.
\end{align}

\section{Partial-conditioning scheme}
\label{App:partial-conditioning scheme}
The ``partial-conditioning'' scheme decomposes the probability Eq.\;\ref{Ptauij} into two parts:
\begin{align}\label{partial_con}
P(\tau, i, j) = P_1(\tau) \cdot P_2(i,j|\tau)\,,
\end{align}
where $P_1(\tau)$ is the same as the probability density that we have defined in Eq.\,\ref{Eq:prob_full}, and $P_2(i,j|\tau)$ is the conditional probability that PBHs $i$ and $j$ will participate the next merger given that the merger occurs at time $t+\tau$. These two probabilities can be written as 
\begin{align}
P_1(\tau) = \frac{\mathcal{K}_0}{V} \exp \left( -\frac{\mathcal{K}_0}{V} \tau \right), \quad P_2(ij|\tau) =  \frac{\mathcal{K}_{ij}}{\mathcal{K}_0},
\end{align}
with the normalization factor $\mathcal{K}_0$ defined as
\begin{align}
\mathcal{K}_0 = \sum_{i=1}^{N-1} \mathcal{K}_i = \sum_{i=1}^{N-1} \sum_{j=i+1}^{N} \mathcal{K}_{ij}\,,
\end{align}
and it is easy to check that $P_1(\tau)$ and $P_2(ij|\tau)$ are properly normalized. As in the full-conditioning scheme, we generate $\tau={V}/{\mathcal{K}_0} \ln \left( {1}/{r_1} \right)$ with a uniformly distributed random number $r_1 \in [0, 1]$.
	
In the partial-conditioning scheme, the pair \(i\) and \(j\) should be generated together, which is best achieved using the acceptance-rejection method (see Appendix\;\ref{Append:ARS} for a basic introduction). Here we employ the two-variable acceptance-rejection method. In this method, we first determine the upper bound \(\mathcal{K}_{\text{max}}\) of the current merger kernel, ensuring that \(\mathcal{K}_{ij} \leq \mathcal{K}_{\text{max}}\) for all \( i = 1, \dots, N-1 \) and \( j = i+1, \dots, N \). This upper bound is used as a constant envelope function. Next, a pair \((i,j)\) is drawn from the uniform distribution over the \(i\)--\(j\) plane of \(P_2(ij|\tau)\). Since \(i \neq j\), the sampling domain is limited to a triangular region. A random number \(r_2 \in [0,1]\) is then generated and the pair is accepted based on the ratio of its merger kernel to the upper bound
\begin{align}
\left\{
\begin{array}{l}
\mathcal{K}_{ij}/\mathcal{K}_{\text{max}}< r_2\,,\quad \textbf{reject $(i,j)$}\\\\
\mathcal{K}_{ij}/\mathcal{K}_{\text{max}}\geq r_2\,,\quad \textbf{accept $(i,j)$}
\end{array}
\right.
\end{align} 

\subsubsection*{The basic simulation algorithm}
After completing the sampling process, the simulation algorithm proceeds as follows (we solve the dimensionless Smoluchowski equation):
\begin{itemize}
\item Step 0: Set the initial time $\tilde{t}_0 = 0$ and the initial number density $\tilde{n}_0 = 1$. Initialize the masses of $N$ PBHs as $\textbf{mass} = [\texttt{m}_1, \texttt{m}_2, ..., \texttt{m}_N]$. For a monochromatic mass distribution, set $\textbf{mass} = [m_0, m_0, ..., m_0]$, which simplifies to $\textbf{mass} = [1, 1, ..., 1]$ in the dimensionless unit. Set $\texttt{m}_i = 0$ for absent PBHs and initialize the maximal merger kernel as $\tilde{\mathcal{K}}_{\text{max}}=\tilde{\mathcal{K}}_{11}$. 

\item Step 1: Generate a random triplet $(\tilde{\tau}, i, j)$ using the partial-conditioning scheme with the acceptance-rejection method. Concretely, compare the merger kernel $\tilde{\mathcal{K}}_{ij}$ to the current maximum value: if $\tilde{\mathcal{K}}_{ij}/\tilde{\mathcal{K}}_{\text{max}} > 1$, update $\tilde{\mathcal{K}}_{ij}\rightarrow\tilde{\mathcal{K}}_{\text{max}}$ and redo the selection. Otherwise, generate a random number $r_2 \in [0,1]$ and accept the pair if $\tilde{\mathcal{K}}_{ij}/\tilde{\mathcal{K}}_{\text{max}} \geq r_2$; otherwise, reject it and select a new pair.  

\item Step 2: For an accepted PBH pair $(i, j)$, update their masses as $\texttt{m}_i + \texttt{m}_j\to\texttt{m}_j$ and set $\texttt{m}_i = 0$. Advance the time variable by $\tilde{t} \to \tilde{t} + \tilde{\tau}$, where  $ \tilde{\tau} = \left({N}/{\tilde{n}_0 \tilde{\mathcal{K}}_0}\right)\ln \left( {1}/{r_1} \right)$.

\item Step 3: Stop the calculation if the total number of PBHs is $1$ or if $\tilde{t} \geq \tilde{t}_f$ with \(\tilde{t}_f\) being the time when the simulation is forced to terminate. Otherwise, return to Step 1.
\end{itemize}

The advantage of acceptance-rejection method is that it does not require explicit knowledge of the target probability density. Its implementation is straightforward and it has modest memory requirements. However, there are several issues.

The first is the choice of envelope function. Because it must upper-bound the target PDF, it needs updating after each newly generated PBH. For simplicity one often uses the PDF’s maximum as a constant (uniform) envelope, which makes sampling straightforward. However, there are drawbacks for a uniform envelope. 
As to obtain the maximum value of the merger rate at each step, the merger rate for every pair of PBHs needs to be computed, which requires $N^2$ calculations and is computationally expensive. To address this, solutions such as \emph{fast acceptance-rejection scheme}\;\cite{wei_fast_2013,wei_gpubased_2013,xu_fast_2014} have been introduced. Fortunately, in our case, since the initial PBH distribution is monochromatic, this issue can be effectively solved. Specifically, we first set an initial maximum value and at each step we compare the merger kernel $\tilde{\mathcal{K}}_{ij}$ of the selected PBHs with the maximum value used in that step, choosing the larger of the two. This ensures that the envelope function at each step still encompasses the entire probability distribution function. 

Although the updating issue can be addressed in our case, the computational efficiency of the acceptance-rejection method remains comparatively low due to its high rejection rate when the merger rate exhibits a large hierarchy, which substantially degrades performance.
The efficiency of sampling,
\begin{align}
\mathcal{E}=\frac{2\tilde{\mathcal{K}}_0}{N(N-1)\tilde{\mathcal{K}}_{\text{max}}}\,,
\end{align}
quantifies the average number of pair selections required to obtain an accepted pair. Additionally, \(\mathcal{E}\) reflects the deviation of the probability distribution from uniformity.
For instance, the merger kernel for \(\texttt{m}_i =\texttt{m}_j = 1\) and \(\texttt{m}_i = \texttt{m}_j = 100\) can differ by five orders of magnitude, while the difference compared to \(\texttt{m}_i = \texttt{m}_j = 10^4\) can reach eleven orders of magnitude—the uniform distribution used as the envelope must be adjusted to accommodate the largest merger kernel. Consequently, for points in the \(i\)--\(j\) plane corresponding to smaller-mass black holes, the imposed upper bound significantly exceeds their \(\tilde{\mathcal{K}}_{ij}\) values. Given the large discrepancies between \(\tilde{\mathcal{K}}_{\text{max}}\) and \(\tilde{\mathcal{K}}_{ij}\), the value of \(\mathcal{E}\) becomes very small, rendering the partial-conditioning scheme slow and inefficient.

To address this issue, several alternatives have been proposed, including \emph{adaptive acceptance-rejection sampling}\;\cite{gilks1992adaptive,thomas_non-uniform_2007}, \emph{ratio-of-uniforms method}\;\cite{kinderman1977computer}, and \emph{adaptive rejection Metropolis sampling}\;\cite{gilks1995adaptive,meyer2008adaptive}. However, these methods are not directly applicable to our problem. First, our model involves discrete variables, so each algorithm would require a discrete adaptation whose validity must be examined for the present setting. Second, identifying a smooth envelope that substantially outperforms a uniform proposal is extremely challenging because merger kernels span many orders of magnitude across masses. Consequently, the probability surface in the \(i\)\(-\)\(j\) plane contains numerous narrow, non-smooth peaks and irregular spikes; these features remain problematic even after variable transformations such as those used by the ratio-of-uniforms approach. Third, because the target is not log-concave, the standard adaptive acceptance-rejection cannot be applied directly. Although the adaptive rejection Metropolis sampling can relax the log-concavity requirement via a Metropolis correction, it still assumes a continuous proposal space and does not avoid the difficulties introduced by discreteness and sharply peaked structure. Finally, the probability function is time dependent, so any efficient envelope must be updated continuously, which further increases computational cost.

\section{The optimization and improvement of efficiency}
\label{App:optimization}
As discussed in Section\;\ref{Sec:Monte_Carlo}, we have supplemented the \emph{unique-mass-mapping method} and \emph{bookkeeping technique} to optimize our simulations. In the following, we elaborate them in detail.

We develop a \emph{unique-mass-mapping} to reduce memory use and improve computational efficiency. For initially calculating $\tilde{\mathcal{K}}_i$, to reduce memory usage and calculation time, we do not have to store the merger kernels of all pairs in a matrix (such a matrix would exceed 64\,GB for $N=10^5$) as in the mass array, masses of the PBHs are not totally different from each other. Especially for all PBHs to initially be monochromatic in mass, we only need to calculate the merger rate of unique masses in the array. 

\begin{figure}[t!]
\centering
\includegraphics[width=0.8\textwidth]{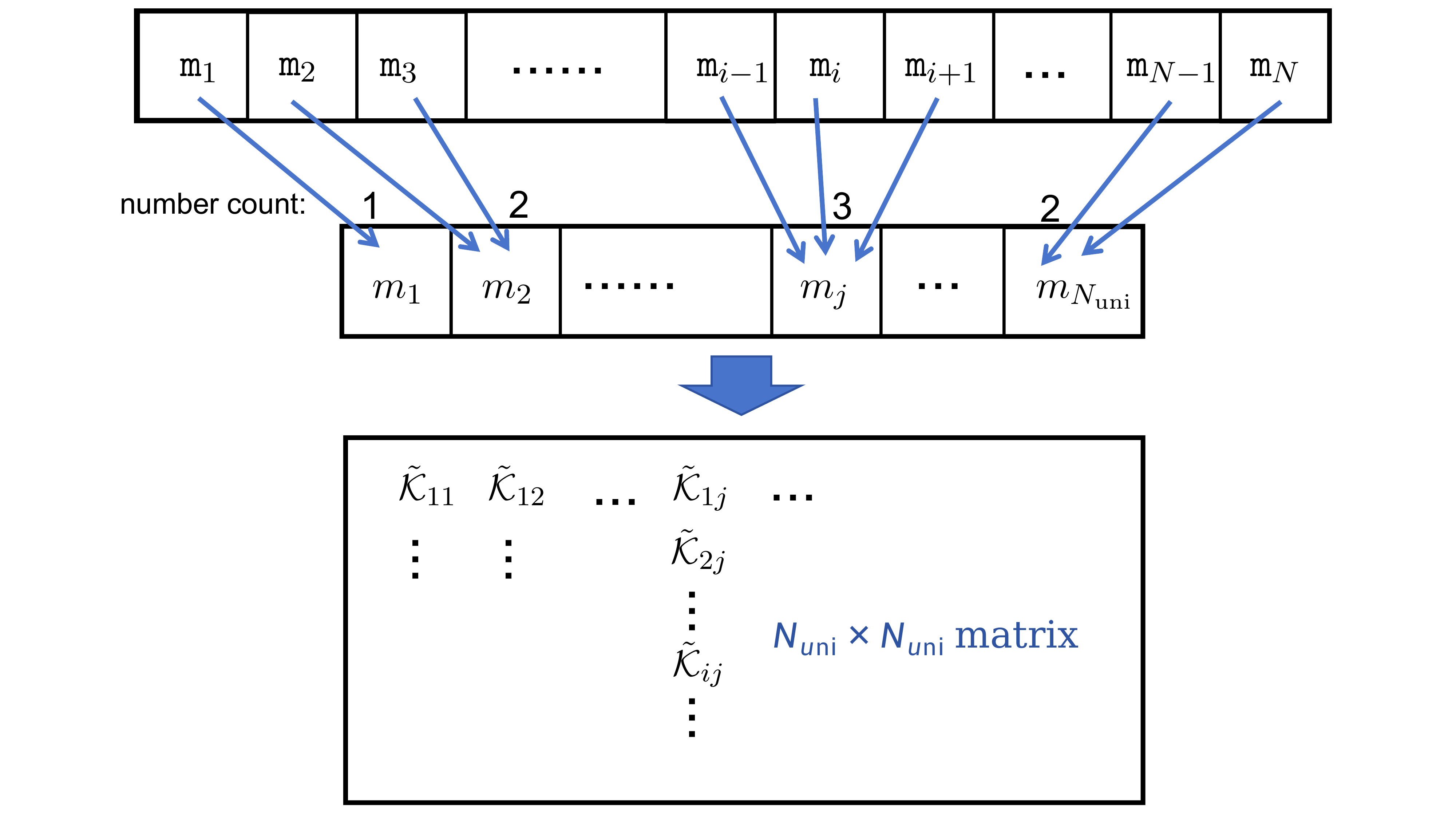}
\caption{Schematic plot of the two mappings}\label{Fig:mapping_schematic}
\end{figure}

We can establish two mappings, one from the mass array to unique masses and another from unique masses to merger kernels, see Fig.\,\ref{Fig:mapping_schematic}.  At each step, we only need to compute the impact caused by the changing pairs, updating the unique masses, the mapping vector from the mass array to unique masses, and the dictionary storing the merger kernels between unique masses. This reduces the storage from an $N^2$ matrix to a mapping vector of length $N$ from the mass array to the unique array, a unique-mass vector of length $N_{\rm uni}$ much less than $N$, and a dictionary of merger kernels corresponding to unique masses of length $\sim N_{\rm uni}^2$. Additionally, the merger kernel is symmetric in PBH masses. This reduces computation and memory storage by half. 

Moreover, we can reduce the calculations using the \emph{bookkeeping technique} that we just update the information of the pair involved in the calculation while keeping others unchanged. The full-conditioning method lies in calculating $\tilde{\mathcal{K}}_i=\sum_{j=i+1}^N \tilde{\mathcal{K}}_{ij}$, and each sampling only involves one pair of PBHs. After sampling, their masses are updated to $\texttt{m}_j=\texttt{m}_j+\texttt{m}_i\,,\; \texttt{m}_i=0$ as the occurrence of a merger. In the computation, we can thus retain the information of unmerged PBHs in the mass array and only update the information of PBHs participating in the merger. Specifically, after calculating each pair of PBHs involved in the merger, we apply the following procedure for the other PBHs: For PBHs with indices in the range of $k<i<j$, their corresponding $\tilde{\mathcal{K}}_i$ cumulative sum is updated by subtracting the pre-merger values of $\tilde{\mathcal{K}}(\texttt{m}_k, \texttt{m}_i) + \tilde{\mathcal{K}}(\texttt{m}_k, \texttt{m}_j) $ and adding the post-merger value $\tilde{\mathcal{K}}(\texttt{m}_k, \texttt{m}_j)$; for PBHs with indices in the range of $i < k < j$, at this point $\tilde{\mathcal{K}}_k$ is independent of $i$, so we only subtract the pre-merger value $\tilde{\mathcal{K}}(\texttt{m}_k, \texttt{m}_j)$ and add the post-merger value $\tilde{\mathcal{K}}(\texttt{m}_k, \texttt{m}_j)$; for $k > j$, no update is needed. Then, $\tilde{\mathcal{K}}_i$ cumulative sum is updated for $i$ and $j$: since $\texttt{m}_i = 0$, $ \tilde{\mathcal{K}}_i = 0$, and for $j$, $\tilde{\mathcal{K}}_j$ needs to be recalculated. 
		
Similarly for updating unique masses, mapping from unique masses to merger rates and mapping from masses to unique masses, we just focus on the pair $(i,j)$ involved in the merger. We first handle the pairs before the merger, and remove their information from the unique masses. For indices $i$ and $j$, this is divided into two cases: the first case is when the mass counts of the element corresponding to $\texttt{m}_{i(j)}$ in the unique masses is $1$, which means that after removal, this mass will disappear from the unique masses. We will remove it, while ensuring that the mapping from masses to unique masses remains unchanged for the other parts. Specifically, for the masses corresponding to indices greater than the indices of $i(j)$ in the unique masses, we need to reduce their unique masses indices by one. We can first operate on $i$ and then on $j$. The second case is when the mass count of the element corresponding to $\texttt{m}_{i(j)}$ in the unique masses is greater than 1. In this case, no operation is required for the unique masses or the mapping from masses to unique masses; we only need to reduce the mass counts of the corresponding mass by one. After removing the information of $i$ and $j$ before the merger, we need to update the new information corresponding to $i$ and $j$ after the merger into the vector. First, for $j$, we consider two situations: Whether or not $j$ is in the unique masses. If it is, we leave the list of unique masses and the mappings for the other masses unchanged; we simply add the mass count corresponding to the mass value of $j$ after merger by one and update $j$ to the corresponding unique-mass mapping. If not, we add $j$ to the unique masses, record the corresponding mass counts as 1, and update the mapping from $j$ from masses to unique masses. While since its mass is zero for $i$, the mapping is identically zero.

Also, we have implemented a basic parallelization scheme in Appendix\;\ref{App:CPU_time}. These optimizations significantly improve simulation efficiency: for \(N=10^4\) the runtime is reduced to \(\sim 10\,\mathrm{s}\); for \(N=10^5\), to \(<10^3\,\mathrm{s}\); and for \(N=10^6\), to \(<10^5\,\mathrm{s}\).  To assess the computational efficiency of different sampling schemes—namely, partial- and full-conditioning—we compute the corresponding elapsed CPU runtimes across a wide range of PBH number \(N=N_{\rm cl}\) and present the fitted scaling relations in \(\log_{10}\) space and the result is explicitly shown in Appendix\;\ref{App:CPU_time}.

\section{CPU times for partial- and full-conditioning schemes}
\label{App:CPU_time}
\paragraph{Numba‑Accelerated JIT Compilation and Parallelization} We also implemented several fundamental and straightforward methods to enhance computational performance. Our codes are written in \texttt{Python}, and we leverage Numba’s \texttt{@njit} decorator to compile core computational kernels into optimized machine code, dramatically reducing \texttt{Python}’s interpreter overhead and making tight loops run at near-\texttt{C/C++} speeds. 

\begin{figure}[H]
\centering
\includegraphics[width=0.8\textwidth]{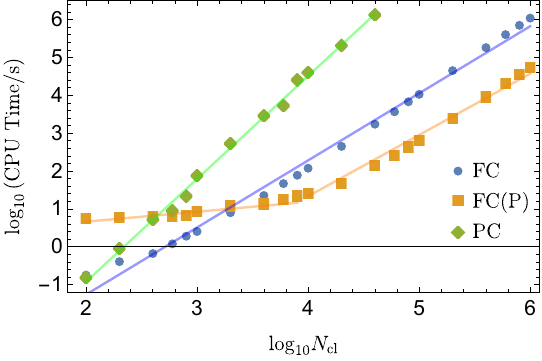}
\caption{CPU times as a function of cluster size $N_{\rm cl}$ for FC, FC(P) and PC schemes}\label{Fig:CPU_timescale}
\end{figure}

Fig.\,\ref{Fig:CPU_timescale} compares the CPU time for both partial‑conditioning (PC) and full‑conditioning (FC) schemes over a cluster size ranging from $N_{\rm cl}=10^2$ to $10^6$ PBHs. The figure displays CPU times for two schemes—single‑GPU implementations in pure \texttt{Python}, denoted “PC” (partial‑conditioning) and “FC” (full‑conditioning)—as well as for the full‑conditioning scheme accelerated with \texttt{@njit} and parallel execution, denoted “FC(P).” The full‑conditioning scheme with discrete inversion transformation method substantially accelerates computation, outperforms the partial‑conditioning scheme with acceptance-rejection method, especially for large $N_{\rm cl}$, where the steep merger‑rate hierarchy renders direct acceptance-rejection sampling extremely inefficient. Moreover, \texttt{@njit} compilation with parallel execution further reduces CPU time nearly an order of magnitude. 

To quantify the trend of efficiency for various cluster sizes $N_{\rm cl}$, we fit the CPU‑time data. For single‑CPU, the PC- and FC-CPU times scale linearly in $\log_{10}$ space, whereas the optimized FC(P)-CPU time requires a two‑segment linear fit:
\begin{align}
&\text{FC}:\, \log_{10}\left(\text{CPU Time}\right)=-4.82+1.77\log_{10}N_{\rm cl}\;.\\
\notag\\
&\text{FC(P)}:\, \log_{10}\left(\text{CPU Time}\right)=A+B\log_{10}N_{\rm cl}\,,\notag\\
&\left\{
\begin{array}{ll}
A=0.13\,,\quad B=0.26\,,&  10^2\leq N_{\rm cl} \leq 6\times10^3\,, \\
A=-5.15\,,\quad B=1.62\,,& 6\times10^3\leq N_{\rm cl}\leq10^6\;.
\end{array}
\right.\\
\notag\\
&\text{PC}:\, \log_{10}\left(\text{CPU Time}\right)=-6.38+2.72\log_{10}N_{\rm cl}\;.
\end{align}

The results show that the FC scheme displays a shallower linear scaling of CPU time with cluster size than the PC scheme, corresponding to a CPU‑time reduction on the order of \(\mathcal{O}\bigl(10^{\log_{10}N_{\rm cl}-2}\bigr)\) at fixed \(N_{\rm cl}\). For the FC(P) case, at large \(N_{\rm cl}\) the scaling behavior is nearly identical to the non‑optimized implementation, since the underlying algorithmic complexity remains. However, the absolute CPU time is lowered by roughly one order of magnitude. For small \(N_{\rm cl}\), the computation times scale shallower, but the absolute values ($N_{\rm cl}\leq 2\times 10^3$) are slightly higher, as upfront compilation overhead and thread‑management costs dominate. For larger \(N_{\rm cl}\), where parallel overheads become subdominant, the optimized implementation again outperforms the non‑optimized version. 

\bibliographystyle{JHEP}
\normalem
\bibliography{smol}

\end{document}